\documentclass[11pt]{article}

 \csname
@addtoreset\endcsname{equation}{section}

\def\a{\alpha}  
\def\b{\beta}   \def\ub{{\underline \b}}
\def\c{\gamma} 
\def\C{\Gamma}
\def\d{\delta} 

\def\e{\epsilon} 

\def\f{\phi}
\def\F{\Phi}

\def\k{\kappa}

\def\m{\mu}
\def\n{\nu}
\def\r{\rho}
\def\s{\sigma}
\def\S{\Sigma}
\def\t{\tau}
\def\th{\theta}

\def\y{\eta}

\def\O{\Omega}

\def\cA{{\cal A}}
\def\cR{{\cal R}}
\def\cS{{\cal S}}
\def\cP{{\cal G}}

\def\cI{{\cal G}'}

\def\cA{{\cal A}}

\def\yb{{\bar y}}

\def\vb{{\bar v}}
\def\ub{{\bar u}}

\let\la=\label
\let\bm=\bibitem
\def\nn{\nonumber}
\newcommand{\eq}[1]{(\ref{#1})}

\newcommand{\w}[1]{\\[0.#1cm]}
\def\eqs#1#2{(\ref{#1}-\ref{#2})}

\def\be{\begin{equation}}
\def\ee{\end{equation}}
\def\bea{\begin{eqnarray}}
\def\eea{\end{eqnarray}}
\def\ba{\begin{array}}
\def\ea{\end{array}}

\def\se{\;\;=\;\;}

\def\mx#1#2#3#4{\left#1\begin{array}{#2} #3 \end{array}\right#4}

\def\ft#1#2{{\textstyle{{\scriptstyle #1}
\over {\scriptstyle #2}}}}

\def\ket#1{|#1\rangle}

\newcommand{\hoch}[1]{$\, ^{#1}$}

\newcommand{\tamphys}{\it\small Center for Theoretical Physics, Texas
A\&M University, College Station, TX 77843, USA}

\newcommand{\groningen}{\it\small Institute for Theoretical Physics,
Nijenborgh 4, 9747 AG Groningen,The Netherlands}

\newcommand{\auth}{\large E. Sezgin\hoch{\dagger}and P. Sundell\hoch{\star} }

\begin{document}

\hfill{CTP-TAMU-13/01}

\hfill{UG-01-27}

\hfill{hep-th/0105001}

\hfill{\today}

\vspace{20pt}

\begin{center}

%%%%%%%%%%%%%%%%%%%%%%%%%%%%%%%%%%%%%%%%%%%%%%%%%%%%%%%%%%%%%%

{\Large \bf Doubletons and 5D Higher Spin Gauge Theory }

%%%%%%%%%%%%%%%%%%%%%%%%%%%%%%%%%%%%%%%%%%%%%%%%%%%%%%%%%%%%%

\vspace{20pt}

\auth

\vspace{15pt}

\begin{itemize}

\item[$^\dagger$] \tamphys

\item[$^\star$] \groningen

\end{itemize}

\vspace{30pt}

{\bf Abstract}

\end{center}

We use Grassmann even spinor oscillators to construct a bosonic
higher spin extension $hs(2,2)$ of the five-dimensional anti-de
Sitter algebra $SU(2,2)$, and show that the gauging of $hs(2,2)$
gives rise to a spectrum $\cS$ of physical massless fields with
spin $s=0,2,4,...$ that is a UIR of $hs(2,2)$. In addition to a
master gauge field which contains the massless $s=2,4,..$ fields,
we construct a scalar master field containing the massless $s=0$
field, the generalized Weyl tensors and their derivatives. We give
the appropriate linearized constraint on this master scalar field,
which together with a linearized curvature constraint produces the
correct linearized field equations. A crucial step in the
construction of the theory is the identification of a central
generator $K$ which is eliminated by means of a coset
construction. Its charge vanishes in the spectrum $\cS$, which is
the symmetric product of two spin zero doubletons. We expect our
results to pave the way for constructing an interacting theory
whose curvature expansion is dual to a CFT based on higher spin
currents formed out of free doubletons in the large N limit. Thus,
extending a recent proposal of Sundborg (hep-th/0103247), we
conjecture that the $hs(2,2)$ gauge theory describes a truncation
of the bosonic massless sector of tensionless Type IIB string
theory on $AdS_5\times S^5$ for large $N$. This implies AdS/CFT
correspondence in a parameter regime where both boundary and bulk
theories are perturbative.

\pagebreak

\setcounter{page}{1}

%%%%%%%%%%%%%%%%%%%%%%%%%%%%%%%%%%%%%%%%%%%%%%%%%%%%%%%%%%%%%%%%%%%%%

\section{Introduction}

%%%%%%%%%%%%%%%%%%%%%%%%%%%%%%%%%%%%%%%%%%%%%%%%%%%%%%%%%%%%%%%%%%%%%

Motivations for studying higher spin fields have varied in time.
To begin with, ``they are there'', in the sense that there exist
higher spin representations of the Poincar\'e group and therefore
it is natural to seek field theories which would describe
particles that carry these representations. In fact, already in
1939, Fierz and Pauli \cite{fp} studied the field equations for
massive higher spin fields. They studied the free field equations,
and while potential difficulties in constructing their
interactions were recognized, the real difficulties became more
transparent much later. The Fierz-Pauli type equations in flat
spacetime were developed further in 1974 \cite{sh} and their
massless limits were obtained in 1978 \cite{f1,ff1}.

Difficulties in constructing the interaction of higher spin fields
were better understood by the early eighties, both in the S-matrix
\cite{gpn,ww} and field theoretic \cite{ad,b,dWF} approaches.
These studies, which led to certain no-go theorems, made certain
assumptions though, which turn out to be too restrictive as was
discovered later. Among the assumptions made were Lorentz
invariance (thus, neglecting the possibility of anti-de Sitter
invariance, for example) and the fact that one higher spin field
at a time was considered (thus, leaving open the consequences of
introducing infinitely many higher spin fields).

Interest in a search for consistent interactions of massless
higher spin fields received a boost with the discovery of
supergravity in mid-seventies. Among the reasons for the renewed
interest in the subject were: a) to better understand the
uniqueness of supergravity theory; b) to search for supergravities
with higher ($N>8$) extended supersymmetry which would involve
larger Yang-Mills gauge symmetries with better grand unification
chances; c) the possibility of a better quantum behaviour by the
inclusion of higher spin gauge fields; and d) from a purely
theoretical point of view, to develop a deeper understanding of
gauge theories that goes beyond Yang-Mills and include massless
fields of arbitrary spin. Most of the attempts made in these
directions, some of which are mentioned briefly above,
essentially led to negative results. Nonetheless, an interesting
development took place in 1978 and the massless higher spin gauge
theory problem was shifted to what appeared to be a more
complicated setting, namely anti-de Sitter space. In fact, this
shift turned out to be crucial for the subsequent breakthroughs
that took place in the development of higher spin gauge theory,
as we shall explain briefly below.

In 1978, Flato and Fronsdal \cite{ff} established that the
symmetric product of two ultra-short representations of the
anti-de Sitter group in four dimensions, known as the singletons,
yields an infinite tower of higher spin massless representations.
Motivated by this result, the free field equation for massless
fields of arbitrary spin in $AdS_4$ were constructed in 1978
\cite{f2,ff2}. Moreover, it was suggested by Fronsdal \cite{f2}
that ``a theory of interacting singletons will provide an example
of interactions between massless fields with higher spins''. We
will come back to this point later.

Nearly a decade later, in 1987, Fradkin and Vasiliev \cite{fv,fv0}
made an important dent in the problem of interacting higher spin
gauge theory. They showed that the gravitational interaction of
massless higher spin fields does exist after all, provided that
the construction is based on an infinite dimensional extension of
the $AdS_4$ algebra and that the interaction is expanded around an
$AdS_4$ background. One may argue that the key to this development
is the recognition of the importance of a suitable choice of
higher spin symmetry algebra. It is intuitively clear that once a
generator with spin higher than two is introduced, insisting on a
Lie algebra, its closure will require the introduction of an
infinite set of higher spin generators as well. To see this, it is
sufficient to consider the AdS generators as bilinears of suitably
chosen oscillators with natural commutation rules, and the higher
spin generators as polynomials of higher than quadratic order in
these oscillators. What is more surprising, at least at first
sight, is the the fact that the flat space limit cannot be taken
in the interactions involving higher spin fields. On the other
hand, this is how the theory manages to circumvent the no-go
theorems of \cite{gpn,ww} which were based on Minkowskian S-matrix
considerations.

Until the late eighties, the massless higher spin gauge theories
were mainly being considered in their own right, though
undoubtedly a great deal of motivation must have been gathered
from the by then well-established conviction that higher spin
gauge theories do exist and as such should necessarily have to
have a bearing on unified theories involving gravity.
Nevertheless, no particular significance was yet attached to the
remarkable connection between $AdS_4$ singletons and massless
higher spin fields discovered nearly a decade ago \cite{ff}.
Moreover, no connection with string theory or any theory of
extended objects was made yet despite the fact that the theory
contains an infinite set of fields of ever increasing spin which
is, in spirit, reminiscent of the spectrum of string theory. This
situation changed soon after the discovery of the supermembrane in
1987 \cite{bst}. A great deal of attention was given to the fact
that its local fermionic symmetries require $D=11$ supergravity
equations of motion to be satisfied \cite{bst}. This led to the
suggestion that supergravity in $D=11$ could be considered as the
low energy limit of a supermembrane theory, though, as we know,
this issue is still not entirely well understood. In any event,
the discovery of the supermembrane-supergravity connection
motivated the study of the Kaluza-Klein vacua of $D=11$
supergravity as vacua for the supermembrane, and special attention
was given to the $AdS_4\times S^7$ solution.

Soon after the discovery of the eleven dimensional supermembrane,
it was suggested  in \cite{duff1} that the singletons could play a
role in its description. Subsequently, it was conjectured in
\cite{bd1} and \cite{nst} that a whole class of
singleton/doubleton field theories constructed on the boundary of
certain $AdS_{p+1}$ spaces described super $p$-branes propagating
on $AdS_{p+1}\times S^{N}$ which existed for certain $p$ and $N$
as supersymmetric Kaluza-Klein backgrounds of a class of
supergravity theories. Yet, the $AdS_7\times S^4$ compactification
of $D=11$ supergravity and the $AdS_5\times S^5$ compactification
of Type IIB supergravity in $D=10$ were inexplicably overlooked;
should they have also been considered, they would have pointed to
the existence of the five-brane in $D=11$ and the Type IIB
three-brane in $D=10$, several years before their actual
discovery.

Once the idea of the eleven dimensional supermembrane on
$AdS_4\times S^7$ being described by the singletons was
entertained, it was natural to consider the possibility of $AdS_4$
higher spin fields arising in the spectrum of the supermembrane.
In 1988, Bergshoeff, Salam, Sezgin and Tanii \cite{bsst} proposed
that the spectrum of the supermembrane in the $AdS_4\times S^7$
background (treated as a second quantized singleton field theory
in three dimensions) contains the massless higher spin states
contained in the symmetric product of two $N=8$ supersingletons.
These states fill irreps of $OSp(8|4)$ with highest spin
$s_{max}=2,4,6,...$, where the $s_{max}=2$ multiplet corresponds
to the well known gauged $D=4, N=8$ supergravity. It was also
pointed out in \cite{bsst} that the massive multiplets contained
in the products of three or more singletons would appear in the
spectrum. Moreover, it was realized in \cite{bsst} that while the
singleton field theory is free, it will nonetheless yield
interactions in the bulk of $AdS_4$, in analogy with 2D free
conformal field theory being capable of describing interactions in
10D target space. It was also suggested in \cite{bsst} that the
resulting theory in $AdS_4$ could provide a field theoretic
realization of the infinite dimensional higher spin algebras of
the kind considered by Fradkin and Vasiliev \cite{fv}. Using the
remarkable relation between the singletons and spectrum of higher
spin states mentioned earlier, the ``admissible'' higher spin
algebras were, in fact, determined later by Konstein and Vasiliev
\cite{kv2}.

It is interesting that the massless higher representations would
emerge first in the context of  supermembrane in $AdS_4\times S^7$
background, as opposed to string theory, perhaps arising from the
Regge trajectory of massive states at high energies. In fact, in
1991, Fradkin and Linetsky \cite{FL2}, conjectured that ``there
might be some sort of phase transition in string theory at high
energies when the cosmological constant of the Planck order is
induced and an infinite dimensional AdS higher spin gauge symmetry
is restored''. These authors were motivated by the similarity
between the non-analyticity in string tension of the massive
string states and the non-analyticity in cosmological constant of
the higher spin gauge theory. In fact, an attempt was made in
\cite{FL2} to study string theory in AdS target space, with
emphasis on determining the critical value of the cosmological
constant to ensure freedom from worldsheet conformal anomalies.
These authors also suggested in \cite{FL2} the possibility of
massless higher spin fields emerging from the product of two
singleton states. In retrospect, one wonders why string theory in
$AdS_5 \times S^5$ was not considered already then in this
context.

Regardless of the considerations of a possible connection with
strings or membranes, the theory of a consistent, interacting
higher spin gauge theory initiated by Fradkin and Vasiliev
\cite{fv,fv0} was developed further by Vasiliev in a series of
papers. In particular, spin zero and half fields were introduced
to the system within the framework of free differential algebras
\cite{v6}. The need to introduce these matter fields fitted nicely
with the fact that they correspond precisely to the spin zero and
half states that arise in the product of two singletons that carry
the representation of the appropriate admissible higher spin
algebra. The theory was furthermore cast into an elegant
geometrical form in \cite{v7} by extending the higher spin algebra
to include new auxiliary commuting spinorial variables. The
resulting formulation of the theory is a free differential algebra
containing a master gauge field and a master scalar field defined
in an extended spacetime which has the usual four (commuting)
spacetime coordinates as well as a set of non-commutative
Grassmann even spinorial coordinates. The non-commutativity is
defined by a star-product involving non-trivial contractions
between the extended spinor coordinates as well as between these
coordinates and the algebra oscillators. Solving the constraints
in the extended directions and evaluating the remaining
constraints at a subspace isomorphic to the ordinary spacetime
leads to a deformed free differential algebra in the space time
that includes interactions and that gives the correct free
massless higher spin equations upon linearization. As such, the
theory is realized at the level of field equations, but an action
from which these field equations can be derived is not known.

The advances made in 1995 with the emergence of D-branes and
M-theory, while highlighting the importance of branes and the role
of eleven dimensions, did not revive interest in higher spin gauge
theory. In fact, the eleven dimensional membrane, which now was
being referred to as the M2-brane, became one of the many possible
branes that existed in a $p$-brane democracy, that had M5-branes,
several D-branes, and other kinds of branes as well. As for the
AdS background, although it was realized that certain brane
solutions extrapolated between Minkowski spacetime and AdS space
\cite{gt}, the surprisingly powerful consequences of AdS
background were really appreciated first in 1997 with Maldacena's
conjecture \cite{jm} on the correspondence between physics in the
bulk of AdS and conformal field theory on its boundary. As the
main argument for the conjecture is that the AdS physics should
actually be decoupled string theory or M-theory in the near
horizon region of a brane (in some suitable limit), and as the
precise formulation of string/M-theory in AdS spacetimes is still
not under control, most attention has been focused, however, on
the issue of testing a weaker form of the correspondence in the
context of gauged supergravities, which are expected to describe
the low energy limits of string/M-theory in AdS backgrounds.

This development motivated us to revisit the higher spin gauge
theory \cite{ss1,ss2,ss3}. The canonical examples of AdS/CFT
correspondence are the maximally supersymmetric cases of Type IIB
string theory in $AdS_5\times S^5$ background and $M$-theory in
$AdS_4\times S^7$ and $AdS_7\times S^4$ backgrounds. Since the
higher spin gauge theory had been worked out in detail in $AdS_4$
and not yet in $AdS_5$ and $AdS_7$, we naturally studied further
the case of $AdS_4$.

The main focus of our work in \cite{ss1,ss2} was to show how
gauged $D=4,N=8$ supergravity is embedded in Vasiliev's higher
spin gauge theory, and to elucidate the geometrical structure of
these equations. While the embedding has been exhibited at the
level of linearized field equations, interesting mysteries remain
to be solved, as far as the nonlinear embedding is concerned.
Considering the simplest bosonic higher spin gauge theory, various
aspects of a curvature expansion scheme advocated by Vasiliev
\cite{v6} was re-visited in \cite{ss3} but a detailed study of the
nonlinearities in the theory is still lacking. In \cite{ss1,ss2},
aspects of singleton dynamics on the boundary of $AdS_4$ yielding
information on the higher spin gauge theory in the bulk were also
discussed but were not put into a concrete mathematical
foundation.

The case of $AdS_5$ appears to be more suitable, however, for
examining the details of the AdS/CFT duality because the $N>1$
version of the CFT is in principle known for this case, unlike the
cases of $AdS_4$ and $AdS_7$, and large  $N$ is actually required
for the AdS radius to be large compared to the Planck length,
which is a basic requirement for the higher spin curvature
expansion scheme to be reliable \cite{ss3}. Of course, if one
assumes that the 4D and yet to be constructed 7D higher spin gauge
theories are actually contained in M-theory, one could infer the
properties of large $N$ M2 and M5 brane dynamics from the
corresponding higher spin curvature expansions. However, awaiting
such a development, it is natural to focus our attention on the
construction of a higher spin gauge theory in $AdS_5$.

Indeed, in an interesting recent development, the authors of
\cite{su1,su2} have gathered evidence for that the physics of
tensionless Type IIB strings in $AdS_5\times S^5$ background
involves massless higher spin fields. They also sketch a
computational scheme for $N=4$ supersymmetric Yang-Mills theory at
zero coupling 't Hooft coupling and for large $N$, to support
their arguments. This is to be contrasted with the original large
$N$ and large 't Hooft coupling limit of Maldacena \cite{jm}, in
which case, as is well known, strongly coupled Yang-Mills theory
furnishes a holographic description of Type IIB strings with
finite tension. The latter setup has had only a limited scope,
though, when it comes to actually verifying the AdS/CFT
equivalence, due to the lack of computational schemes for the
boundary and bulk theories. Indeed, mostly the implications of the
gauged supergravity (valid at low energies) for the strongly
coupled gauge theory on the boundary have been studied so far.

In the new limit proposed above we expect that computations can be
performed on the higher spin supergravity side using the
above-mentioned curvature expansion technique as well as on the
boundary CFT side using the techniques of \cite{su1,su2}. Hence,
this limit offers a arena for directly verifying the AdS/CFT
conjecture! It remains, however, to find the interactions of the
five-dimensional higher spin gauge theory, which we believe is an
interesting and feasible technical problem. In this paper we have
already taken the first steps in this directions by identifying an
appropriate higher spin algebra in 5D and studying its gauging,
thereby providing what we believe to be an appropriate framework
for the construction of the full theory. Indeed, our results so
far indicate that the theory  has a form similar to the
four-dimensional higher spin gauge theory. We shall return to the
above issues in Sec. 6.

While the higher spin gauge theory is well developed in $D=4$, at
least at the level of writing down full equations of motion in a
concise and geometrical fashion, much less is known in $D>4$. To a
large extent this is due to the fact that the truly universal
principles underlying the known four-dimensional case have not yet
been identified completely. Once these principles are well
understood, a natural strategy would, of course, be to apply them
to any higher dimensions. This is the approach that we will take
here. In fact, such a philosophy was also adopted by Vasiliev in
1990 \cite{v8} who considered some aspects of the problem in
$D=2n$, though the results do not appear to be conclusive. Much
more is known, of course, about higher spin gauge theory in
arbitrary dimensional AdS space at the free level
\cite{vhd1,vhd2,m4,m5,sv}. The reason is that the linearized
theory can be constructed without any knowledge of the underlying
higher spin Lie algebra. One of the main aims of this paper is to
remedy this situation in $D=5$ (see below).

Another approach that has been proposed for studying the higher
spin gauge theory problem in arbitrary dimensional AdS space is to
consider a point particle in a higher spin gauge field background
and to associate the higher spin gauge symmetry with the geometry
of the point particle phase space \cite{as2}. So far, this
approach seems to be rather restrictive and does not seem to make
contact with Vasiliev's  higher spin gauge theory in $D=4$.
However, recently an interesting connection has been made with the
results of \cite{as2} for $AdS_d$, starting from a noncommutative
$Sp(2,R)$ gauge theory with two times in $(d,2)$ dimensions and
then fixing a particular gauge. It is suggested in \cite{bars1}
that some of the difficulties encountered in \cite{as2} may be
overcome in their approach. Moreover, they also make a connection,
as in \cite{su1,su2}, between the higher spin gauge theory and the
zero-tension limit of string theory.

The approach of \cite{bars1} is certainly an interesting one to
pursue. However, it is by no means clear at present how to
reproduce even the existing  higher spin gauge theory of Vasiliev
in $AdS_4$ in that approach. Therefore, being armed with the
knowledge we have gained from Vasiliev's theory in $AdS_4$, we
choose in this paper the ``building up approach" in which we try
to carry over the basic principles of the construction that works
in $D=4$ to higher dimensions, beginning with $AdS_5$. In spirit,
this is similar to search for a superspace formulation of
supergravity field equations in terms of torsion and curvature
constraints. One can then consider a superbrane action in curved
superspace, whose $\kappa$ symmetry would require that these
constraints are satisfied. Historically, supergravity-brane
connections have arisen in this manner. Here too, after
establishing a higher spin gauge theory in terms of Vasiliev style
constraints, one can search for a brane theory which will require
those constraints. It would be rather amusing if one could
discover the unknown higher spin gauge theory constraints by
guessing an appropriate brane action to begin with and then
requiring a suitable local symmetry.

The first principle in Vasiliev's approach to higher spin gauge
theory in AdS background is to identify an appropriate higher spin
gauge symmetry algebra. In doing so, we expect the fundamental
representations of the AdS algebras which are known as singletons
or doubletons to play a crucial role, just as they do in $D=4$
\cite{bsst,kv2}. In this paper, we consider the bosonic higher in
gauge theory in $D=5$ and we find the suitable higher spin
algebra, starting from the doubleton representations of the
$AdS_5$ group $SO(4,2)$ \cite{g1}. As in $D=4$, it is obtained by
using Grassmann even spinor variables, that are complex Dirac
spinors of the spin extension $SU(2,2)$ of $SO(4,2)$, and we have
therefore named it $hs(2,2)$. The next step is to introduce
suitable master fields which form representations of $hs(2,2)$ and
to define the associated curvatures and gauge transformations.
Then, one has to find the suitable constraints on the curvatures
such that their solution will give rise to certain auxiliary
fields and dynamical fields, and moreover the latter ones should
correspond to the spectrum of massless higher spin representations
that arise in the tensor product of two doubletons. To ensure that
the basic setup is right, we then check the resulting linearized
field equations
\footnote{At the algebraic level, the oscillator algebra offers
competing options for defining the higher spin algebra and the
master scalar fields, which all lead to the same field content. At
the linearized level, the definitions made here reproduces the
correct spectrum.}.

We believe that in this paper we have established a framework for
introducing interactions by deforming the linearized system. For a
description of the deformation story, which is a crucial
ingredient of the interacting theory as we understand it
presently, see for example the reviews \cite{vr2,ss2}. We hope to
return to the deformation problem in $D=5$ in the future, by
making use of the formalism established here. As for the
generalization to the case of higher spin supergravity theory with
32 real supersymmetries in $D=5$, we have carried out essentially
the same steps as in this paper, and those results will appear
elsewhere \cite{ss4}.

The organization of this paper is as follows: In Sec. 2, we review
the representation theory of $SO(4,2)$, and in particular we
discuss in great detail the doubleton representations and derive
the decomposition formula for the product that underlies the
spectrum. In Sec. 3, we define the bosonic higher spin algebra in
$D=5$, which we call $hs(2,2)$, and derive its massless unitary
representation, which consists states with spins $s=0,2,4,...$. In
Sec. 4, we gauge the algebra $hs(2,2)$ by introducing a master
gauge field, and show that its $SO(4,2)$ field content agrees
with the previous suggestions \cite{vhd1,vhd2} from linearized
analysis. In addition we introduce a master scalar field which
forms a representation of $hs(2,2)$ and it is shown to be
necessary to describe the physical scalar as well as the Weyl
tensors, which are the on-shell non-vanishing components of the
master curvature two-form.  In Sec. 5, we write the linearized
constraints on the curvature two-form and the covariant
derivative of the scalar master field. These are shown to be
integrable and to yield correctly linearized field equations in
$AdS_5$, namely the scalar Klein-Gordon equation and the curvature
constraints on auxiliary and dynamical gauge fields written in
the tensorial basis of \cite{vhd1,vhd2}. In Sec. 6, we summarize
our results and speculate over future directions; in particular on
the prospects of a connection with string/M-theory. A more
detailed discussion of possible connections between higher spin
gauge theories and M-theory will be given elsewhere \cite{ss4}.

%%%%%%%%%%%%%%%%%%%%%%%%%%%%%%%%%%%%%%%%%%%%%%%%%%%%%%%%%%%%%%%%%%%%%

\section{Elements of $SO(4,2)$ representation theory}

%%%%%%%%%%%%%%%%%%%%%%%%%%%%%%%%%%%%%%%%%%%%%%%%%%%%%%%%%%%%%%%%%%%%%

In this section we review some basic elements of the
representation theory of $SO(4,2)$ \cite{g1} that will be needed
for the analysis of the higher spin algebra given in the next
section. We also refine the work of \cite{g1} in the sense that we
compute the multiplicity of the higher spin massless weight spaces
that occur in the decomposition in the product of two spin zero
doubletons. This data is necessary for the application to the
five-dimensional higher spin gauge theory in order to deduce the
precise form of the spectrum in the next section.

The five-dimensional AdS group and the four-dimensional conformal
group are isomorphic to $SO(4,2)$ ($A=0,\dots,3,5,6$):

\be [M_{AB},M_{CD}]=-i(\y_{BC}M_{AD}+\mbox{$3$ more})\
,\la{so42}\ee
where $\y_{AB}={\rm diag}(-1,+1,+1,+1,+1,-1)$. The maximal compact
subgroup of $SO(4,2)$, or rather its spin extension $SU(2,2)$, is
$L^0=S(U(2)\times U(2))=SU(2)_{\rm L}\times SU(2)_{\rm R}\times
U(1)_E$, which can be taken to be represented by the generators
($i=1,2,3$)
\footnote{We define $P_a=-M_{a6}$ ($a=0,...,3,5$). The generator
$P_0=E=-M_{06}$ is the AdS energy in five dimensions. In four
dimensions $E$ is the conformal Hamiltonian, while $D=-M_{56}$ is
the generator of dilatations.}:

\bea SU(2)_{\rm L}&:&
L_i=\ft12(M_{5i}+\ft12\epsilon_{ijk}M_{jk})\ ,\\
SU(2)_{\rm R}&:&
R_i=\ft12(-M_{5i}+\ft12\epsilon_{ijk}M_{jk})\ ,\\
U(1)_E&:& E=-M_{06}\ .
 \eea
The remaining generators of $SO(4,2)$ split into a space $L^+$ of
energy-raising operators and a space $L^-$ of energy-lowering
operators, such that $[E,L^\pm]=\pm L^\pm$,
$[L^+,L^+]=0=[L^-,L^-]$ and $[L^+,L^-]=L^0$. Thus unitary positive
energy representations of $SO(4,2)$ (with the reality condition
$(M_{AB})^{\dagger}=M_{AB}$ implying $(L^+)^{\dagger}=L^-$)
consist of weight spaces $D(j_{\rm L},j_{\rm R};E)$ formed by
acting with $L^+$ on a space of ground states, or lowest weight
states, $\ket{j_{\rm L},j_{\rm R};E}$ which are annihilated by
$L^-$ and form a representation of $L^0$ labeled by $(j_{\rm
L},j_{\rm R};E)$. Such representations can be obtained by taking

\be M_{AB}=\ft12 \yb\star \S_{AB} y=\ft12 \yb \S_{AB} y\
,\la{m}\ee
where the four-component $SO(4,1)$ Dirac spinor $y_\a$ and its
conjugate $\yb^\a$ obey the oscillator algebra
\footnote{The
five-dimensional Dirac matrices $\C_a$ ($a=0\dots5$) obey
$\{\C_a,\C_b\}=2\eta_{ab}$. We define the Dirac conjugate
$\yb^{\a}=(y^{\dagger}i\C^0)^\a$ and the Majorana conjugate
$\yb_\a=\yb^\b C_{\b\a}$. The anti-symmetric conjugation matrix
$C_{\a\b}$ obeys $C_{\a\b}C^{\c\b}=\delta_\a^\c$ and a reality
condition such that $(\bar{\psi}\chi)^{\dagger}=\bar{\chi}\psi$,
where we by definition set
$(\psi\chi)^{\dagger}=\chi^{\dagger}\psi^{\dagger}$. The matrix
$(\C^a C)_{\a\b}$ is anti-symmetric and $(\C^{ab} C)_{\a\b}$ is
symmetric.}

\be [y_\a,\yb^\b]_*=2\d_\a^\b\ ,\quad y_\a\star
\yb^\b=y_\a\yb^\b+\d_\a^\b\ ,\quad \yb^\a\star y_\b=\yb^\a
y_\b-\d_\b^\a\ ,\la{o}\ee
where $\star$ denotes the operator product and $y_\a\yb^\b$ the
Weyl ordered product, and

\be \S_{ab}=-\ft{i}2\C_{ab}\ ,\quad \S_{a6}=-\ft{i}2\C_a\ .\ee
{}From $(\S_{AB})^{\dagger}=-\C^0\S_{AB}\C^0$ it follows that
$(M_{AB})^{\dagger}=M_{AB}$, and $E=\ft14 y^{\dagger}y$ shows that
the energy is positive. The representation content of the
oscillator Hilbert space can be listed by going to the standard
representation of the Dirac matrices:

\be \C^0=i\mx{(}{cc}{1&0\\0&-1}{)}\ ,\quad
\C^i=i\mx{(}{cc}{0&\s^i\\-\s^i&0}{)}\ ,\quad
\C^5=i\C^0\C^1\C^2\C^3=\mx{(}{cc}{0&1\\1&0}{)}\ ,\la{d}\ee
which splits $y_\a$ into the following pair of $U(2)$ invariant
oscillators ($I=1,2$, $P=1,2$):

\be y_\a=\sqrt{2}\mx{(}{c}{a^I\\b_P}{)}\ ,\quad
\yb^\a=\sqrt{2}(-a_I,b^P)\ ,\quad a^I=(a_I)^{\dagger}\ ,\quad
b^P=(b_P)^{\dagger}\ ,\la{ab}\ee

\be [a_I,a^J]_*=\d_I^J\ ,\quad [b_P,b^Q]_*=\d_P^Q\ .\ee
The compact $SO(4,2)$ generators are then given by:

\be
L_i=\ft 12(\s^i)_I{}^J L_J{}^I\ ,\quad
R_i=\ft12(\bar{\s}^i)^P{}_Q R^Q{}_P\ ,
\ee
$$
E=\ft12(a^Ia_I+ b^P
b_P)=\ft12(N_a+N_b+2)\ ,
$$
where $\bar{\s}^i=-(\s^i)^{\star}$ and

\be
L^I{}_J=a^I\star a_J-\ft12 \d_J^I N_a=a^Ia_J-\ft12\delta_I^J
a^Ka_K\ ,
\ee
$$
R^P{}_Q=b^P\star b_Q-\ft12 \d_Q^P N_b=b^Pb_Q-\ft12
\delta_Q^P b^Rb_R\ ,
$$
\be
N_a=a^I\star a_I=a^Ia_I-1\ ,\quad N_b=b^P\star b_P=b^P b_P-1\ .
\la{na}
\ee
The remaining $SO(4,2)$ generators are the energy-lowering
operators $L_{IP}$ and the raising operators $L^{IP}$ given by:

\be L_{IP}=a_I\star b_P=a_I b_P\ ,\quad L^{IP}=a^I\star b^P=a^I
b^P\ ,\ee
satisfying the algebra

\be [L_{IP},L^{JQ}]_*=\d_I^JR^Q{}_P+\d_P^QL^J{}_I+\d_I^J\d_P^QE\ .\ee

Upon letting $\ket{0}$ be the oscillator vacuum obeying

\be
a_I\ket{0}=b_P\ket{0}=0\ ,
\ee
the lowest weight states of the oscillator Hilbert space are given
by

\be \ket{(j,0;j+1)}=a^{I_1}\cdots a^{I_{2j}}\ket{0}\ ,\quad
\ket{(0,j;j+1)}=b^{P_1}\cdots b^{P_{2j}}\ket{0}\ ,\quad
j=0,\ft12,1,\dots \la{lws}\ee
The resulting weight spaces $D(j,0;j+1)$ and $D(0,j;j+1)$ are
known as the $SO(4,2)$ doubleton representations, and correspond
to the mode expansions (in a fixed gauge) of $D=4$ conformal
tensors with $SO(3,1)$ spin $j$. These weight spaces are not
sufficiently large, however, for constructing the mode expansions
of $AdS_5$ tensors. The spectrum of such tensors is contained in
the $N$-fold tensor products $(N>1)$ of the doubleton
representations. Such a tensor product is formally equivalent to
the following oscillator algebra (r,s =1,\dots,N):

\be [a_I(r),a^J(s)]_*=\d_I^J\d_{rs}\ ,\quad
[b_P(r),b^P(s)]_*=\d_P^Q\d_{rs}\ . \ee
The representation of $SO(4,2)$ on the tensor product space is
then given by

\be M_{AB}=\sum_r M_{AB}(r)\ ,\quad M_{AB}(r)=\ft12 \yb(r)\S_{AB}
y(r)\ . \ee

It follows that $E=j_{\rm L}+j_{\rm R}+N$. For $N=2$, that is the
two-fold tensor product, this yields massless representations of
$SO(4,2)$. As a preparation for the next section, we compute the
tensor product of two spin zero doubletons. The weight space
$D(0,0;1)$ is spanned by the states

\be \ket{0}\ ,\quad a^Ib^P\ket{0}\ ,\quad a^Ia^Jb^Pb^Q\ket{0}\
,\quad \cdots\ .\ee
To find the ground states we start from the following general
expansion of a state $\ket{\psi}$ in the tensor product with fixed
energy $E=n+2$:

\bea \ket{\psi}&=&\sum_{k=0}^n\psi^{(k)}_{I(k),P(k);J(n-k),Q(n-k)}
a^{I_1}(1)\cdots a^{I_k}(1) b^{P_1}(1)\cdots b^{P_k}(1)\nn\\&&
\qquad a^{J_1}(2)\cdots a^{J_{n-k}}(2)b^{Q_1}(2)\cdots
b^{Q_{n-k}}(2)\ket{0}\ ,\eea
where we use a condensed notation such that $I(k)=I_1\dots I_k$
denotes $k$ (symmetrized) indices. Acting on this state with the
energy-lowering operators $L_{IP}=a_I(1)b_P(1)+a_I(2)b_P(2)$
should give zero, which amounts to the following set of equations:

\bea
n^2\psi^{(n)}_{IJ(n-1),PQ(n-1)}+\psi^{(n-1)}_{IP;J(n-1),Q(n-1)}&=&0\
,\nn\\
(n-1)^2\psi^{(n-1)}_{KR;IJ(n-2),PQ(n-2)}+4\psi^{(n-2)}_{IK,PR;J(n-2),Q(n-2)}&=&0\
,\nn\\
&\vdots&\\
\psi^{(1)}_{K(n-1),R(n-1);IP}+n^2\psi^{(0)}_{IK(n-1),PR(n-1)}&=&0\
.\nn\eea
{}From the first equation we can solve for $\psi^{(n-1)}$ in terms
of $\psi^{(n)}$, etc. The ground states with $E=n+2$ form an
irreducible representation of $SU(2)_L\times SU(2)_R$ with
quantum numbers $(j_L,j_R)=({n\over 2},{n\over 2})$. The lowest
energy states listed in the order of increasing energy are

\bea
\ket{(0,0;2)}&=&\ket{0}\ ,
\nn\\
\ket{(\ft12,\ft12;3)}&=&\left(a^I(1)b^P(1)-a^I(2)b^P(2)\right)\ket{0}\
,\nn\\
\ket{(1,1;4)}&=&\left(a^I(1)a^J(1)b^P(1)b^Q(1)-4a^{(I}(1)a^{J)}(2)
b^{(P}(1)b^{Q)}(2)\right.\nn\\&&\left.+a^I(2)a^J(2)b^P(2)b^Q(2)\right)\ket{0}
\ ,
\la{hws}\\
&\vdots&\nn\\
\ket{(j,j;2j+2)}&=&\sum_{k=0}^{2j}(-1)^k\left(\!\!{\small
\mx{.}{c}{2j\\k}{.}}\!\!\right)^2 a^{(I_1}(1+\th(k))\cdots
a^{I_{2j})}(1+\th(k+1-2j))
\nn\\
&&\qquad  b^{(P_1}(1+\th(k))\cdots
b^{P_{2j})}(1+\th(k+1-2j))~\ket{0}\ ,
\nn\\
&\vdots&
\nn
\eea
where $\th(x)=0$ if $x\leq 0$ and $\th(x)=1$ if $x>0$. The states
with even spins $s=j_L+j_R=0,2,4,...$ belong to the symmetric
tensor product and the states with odd spins $s=1,3,5,...$ to the
anti-symmetric product:

\bea \left[D(0,0;1)\otimes D(0,0;1)\right]_{\rm
S}&=&\sum_{s=0,2,\dots}D\left({s\over 2},{s\over 2};s+2\right)\
,\la{stp}\\
\left[D(0,0;1)\otimes D(0,0;1)\right]_{\rm
A}&=&\sum_{s=1,3,\dots}D\left({s\over 2},{s\over 2};s+2\right)\ .
\eea
%

%%%%%%%%%%%%%%%%%%%%%%%%%%%%%%%%%%%%%%%%%%%%%%%%%%%%%%%%%%%%%%%%%%%%%

\section{The higher spin algebra $hs(2,2)$ and its spectrum}

%%%%%%%%%%%%%%%%%%%%%%%%%%%%%%%%%%%%%%%%%%%%%%%%%%%%%%%%%%%%%%%%%%%%%

In this section we define a higher spin extension $hs(2,2)$ of
$SO(4,2)$ by the coset $\cP/\cI$, where $\cP$ is a Lie subalgebra
of the algebra $\cA$ of arbitrary polynomials of the oscillators
\eq{o} and $\cI$ is an ideal of $\cP$ generated by a central
element $K$. The basic argument for modding out $K$ is that it is
responsible for a degeneracy in $\cP$ such that $\cP$ contains
infinitely many generators of any given integer spin. The reason
for this is that $K$ has zero spin so that it can be used to build
elements in $\cP$ of arbitrary monomial degree but with fixed
spin. On the other hand, the coset defining $hs(2,2)$ has a finite
number of generators of any given spin. In this section, we also
define the physical spectrum $\cS$ of the five-dimensional higher
spin gauge theory based on $hs(2,2)$. The basic requirement on
$\cS$ is that it must consist of massless $SO(4,2)$ weight spaces
and carry a unitary (irreducible) representation of $hs(2,2)$.
Armed with the algebra $hs(2,2)$ and its massless spectrum $\cS$,
we will gauge $hs(2,2)$ in the next section.

To define the algebra we first define the associative product of
elements in $\cA$, that is Weyl-ordered (regular) functions of the
oscillators $y$ and $\yb$, as follows:

\be F(y,\yb)\star G(y,\yb)=\int d^8ud^8v F(y+u,\yb+\ub)
G(y+v,\yb+\vb)e^{\ub v-\vb u}\ ,\la{st}\ee
where the integration measure is assumed to be normalized such
that $1\star F=F\star 1=F$. This algebra can also be defined by
the following contraction rule:

\bea &&(y_{\a_1}\cdots y_{\a_m} \yb^{\b_1}\cdots \yb^{\b_n})\star
(y_{\c_1}\cdots y_{\c_p} \yb^{\d_1}\cdots \yb^{\d_q})\nn\\
&=& y_{\a_1}\cdots y_{\a_m} \yb^{\b_1}\cdots \yb^{\b_n}
y_{\c_1}\cdots y_{\c_p} \yb^{\d_1}\cdots \yb^{\d_q}\nn\\&+&
mq\delta_{(\a_1}^{(\d_1|}y^{\phantom{\b_1}}_{\a_2}\cdots
y^{\phantom{\a_1}}_{\a_m)} \yb^{\b_1}\cdots \yb^{\b_n}
y^{\phantom{\a_1}}_{\c_1}\cdots y^{\phantom{\a_1}}_{\c_p}
\yb^{|\d_2}\cdots
\yb^{\d_q)}\la{cr}\\&-&np\delta^{(\b_1}_{(\c_1|}y^{\phantom{\a_1}}_{\a_1}\cdots
y^{\phantom{\a_1}}_{\a_m} \yb^{\b_2}\cdots \yb^{\b_n)}
y^{\phantom{\a_1}}_{|\c_2}\cdots y^{\phantom{\a_1}}_{\c_p)}
\yb^{\d_2}\cdots \yb^{\d_q}\nn\\&+&{m(m-1)q(q-1)\over 2}
\delta_{(\a_1\a_2}^{(\d_1\d_2|}y^{\phantom{\a_1}}_{\a_3}\cdots
y^{\phantom{\a_1}}_{\a_m)} \yb^{\b_1}\cdots \yb^{\b_n}
y^{\phantom{\a_1}}_{\c_1}\cdots y^{\phantom{\a_1}}_{\c_p}
\yb^{|\d_3}\cdots \yb^{\d_q)}\nn\\&+&\cdots\ .\nn\eea
A general term obtained by contracting $k$ $y\yb$ pairs and $l$
$\yb y$ pairs is weighted with

\be (-1)^{l}k!l! {m\choose k}{q\choose k}{n\choose l}{p\choose
l}\d^{\d_1\dots \d_k}_{\a_1\dots \a_k}\delta^{\b_1\dots
\b_l}_{\c_1\dots\c_l}\ .\nn\ee
Here we use unit-strength symmetrized Kronecker-deltas defined by
$\delta^{\a_1\dots\a_p}_{\b_1\dots\b_p}=
\delta^{(\a_1}_{(\b_1}\cdots \delta^{\a_p)}_{\b_p)}$. Since $(\ub
v)^{\dagger}=\vb u$ it follows from \eq{st} that:

\be (F\star G)^{\dagger}=G^{\dagger}\star F^{\dagger}\
.\la{dagger}\ee
The following set of linear maps:

\be \t_\y(y_\a)=\y y_\a\ ,\quad \t_\y(\yb_\a)=-\bar{\y}\yb_\a\
,\quad |\y|=1\ ,\ee
act as anti-involutions of $\cA$:

\be \t_\y (F\star G)=\t_\y(G)\star \t_\y(F)\ .\la{ai}\ee
The Lie subalgebra $\cP$ is defined to be the subspace of $\cA$
consisting of elements $F$ obeying

\be \t_\y(F)=-F\ ,\quad (F)^{\dagger}=-F\ ,\la{tc}\ee
and with Lie bracket

\be [F,G]=[F,G]_*=F\star G-G\star F\ .\la{l}\ee
Lie algebras that are similar to $\cP$ have been defined in even
spacetime dimensions by Vasiliev in a slightly different setup
\cite{v8}. The algebra $\cP$ can be expanded in terms of elements
of the form:

\be {1\over (n!)^2}X_{\a_1\dots \a_n,\b_1\dots
\b_n}\yb^{\a_1}\cdots \yb^{\a_n} y^{\b_1}\cdots y^{\b_n}\ ,\quad
n=1,3,5,\dots \la{x}\ee
where the multi-spinor coefficient obeys the following reality
condition:

\be \bar{X}_{\a_1\dots \a_n,\b_1\dots\b_n}\equiv
\bar{X}^{\c_1\dots\c_n,\d_1\dots\d_n}C_{\c_1\a_1}\cdots
C_{\d_n\b_n}=-X_{\b_1\dots\b_n,\a_1\dots \a_n}\ .\la{rc}\ee
Note that the Dirac conjugate multi-spinor
$\bar{X}^{\c_1\dots\c_n,\d_1\dots\d_n}$ is defined by a hermitian
conjugation followed by multiplication with $i\C^0$ of each spinor
index. The elements in \eq{x} with $n=1$ form the subalgebra
$U(2,2)=SU(2,2)\times U(1)_K$, where $U(1)_K$ is generated by the
${\cP}$ central element

\be K=\ft12 \yb y\ ,\quad [K,F]_*=0\ ,\quad F\in \cP\ .\la{k}\ee
{}From \eq{ab} and \eq{na} it follows that

\be K=\ft12(-a^I a_I+b^P b_P)=\ft12(N_b-N_a)\ .\ee
In a unitary irreducible representation of $\cP$ the generator $K$
is given by a real constant. In particular, the algebra $\cP$ can
be represented unitarily on the oscillator Fock space. As
discussed in the previous section, the oscillator Fock space
decomposes into a direct sum of all the doubletons $D(j,0;j+1)$
and $D(0,j;j+1)$ ($j=0,\ft12,1,\ft32,...$). By construction each
doubleton is an irreducible representations of the $U(2,2)$
subalgebra, in which

\bea
K\se  \mx{\{}{l}{\ \ \, j/2 \quad {\rm for}\ \ D(0,j;j+1)\ ,\\ \\
-j/2 \quad {\rm for}\ \ D(j,0;j+1)\ ,}{.}
\la{kj}
\eea
It follows that each doubleton is also an unitary irreducible
representation of $\cP$.

To examine the degeneracy due to the fact that $K$ has spin zero,
we decompose $\cP$ into levels, such that the $\ell$th level is
given by all elements of the form:

\bea &&{1\over(n!)^2}K^{\star k}\star (X^{(k)}_{\a_1\dots
\a_n,\b_1\dots \b_n}\yb^{\a_1}\cdots \yb^{\a_n}y^{\b_1}\cdots
y^{\b_n})\ ,\nn\\&& C^{\c\d}X^{(k)}_{\c\a_1\dots
\a_{n-1},\d\b_1\dots \b_{n-1}}=0\ ,\quad k+n=2\ell+1\ ,\quad
k,n\geq 0\ , \la{tx}\eea
where we emphasize that $X^{(k)}$ is traceless (recall that
$C_{\a\b}$ is the anti-symmetric charge conjugation matrix in
$D=5$), and we use the notation

\be K^{\star k}=\underbrace{K\star \cdots \star K}_{\mbox{$k$
factors}}\ .\ee
For finite polynomials the expansions \eq{x} and \eq{tx} are
equivalent. By making repeated use of \eq{st} \eq{tx} can be
expanded as a leading term ${1\over(n!)^2}K^{k} X^{(k)}_{\a_1\dots
\a_n,\b_1\dots \b_n}\yb^{\a_1}\cdots \yb^{\a_n}y^{\b_1}\cdots
y^{\b_n}$ plus a finite number of terms of lower polynomial
degree; that is, the basis \eq{tx} corresponds to separating out
the $C_{\a\b}$ traces of the basis elements in \eq{x}. Note that
an element in the $\ell$th level is a sum of elements of the form
\eq{x} with $n\leq 2\ell+1$. The basis \eq{tx} yields the
following unique decomposition of $\cP$:

\be \cP=\cP^{(0)}+K\star \cP^{(1)}+K^{\star 2}\star
\cP^{(2)}+\cdots\ . \la{ge} \ee
Since $K$ is central and $\t_\y(K)=-K$, it follows from \eq{tc}
that $\t(X^{(k)})=(-1)^{1+k}X^{(k)}$. Hence $\cP^{(k)}$ is
isomorphic to $\cP^{(0)}$ or $\cP^{(1)}$ for $k$ even or odd,
respectively. We also remark that since $K$ is central and
$K^{\dagger}=K$, the traceless multi-spinors $X^{(k)}$ obey the
reality condition \eq{rc}.

The degeneracy in $\cP$ discussed above due to $K$ having spin
zero suggests that $K$ should be eliminated from the actual higher
spin algebra. The Lie bracket \eq{l} induces a set of brackets
with the following structure:

\be [\cdot,\cdot]\ :\ \cP^{(k_1)}\times \cP^{(k_2)}\ \rightarrow\
\cP^{(k_1+k_2)}+\cP^{(k_1+k_2+1)}+\cdots\ .\la{l2}\ee
Here the direct sum of spaces, which is a finite sum for finite
polynomials, is due to the fact that the Lie bracket \eq{l} does
not preserve the tracelessness condition in \eq{tx}. Thus $K$
cannot be eliminated by simply restricting $\cP$ to $\cP^{(0)}$.
In order to factor out $K$ in a meaningful way we instead let

\be \cI=K\star \cP^{(1)}+K^{\star 2}\star \cP^{(2)}+\cdots \ee
This space forms an ideal in $\cP$, i.e. $[\cP,\cI]_*=\cI$. We can
now define the higher spin algebra $hs(2,2)$ as following coset:

\be hs(2,2)=\cP/\cP'\ .\ee
The elements of $hs(2,2)$ are thus equivalence classes $[F]$ of
elements in $\cP$ defined by

\be [F]=\{G\in \cP\ |\ F-G\in \cI\}\ .\la{hs}\ee
The Lie bracket of $[F]$ and $[G]$ is given by

\be [[F],[G]]=[[F,G]_*]\ .\ee

The spectrum $\cS$ of $D=5$ higher spin gauge theory based on
$hs(2,2)$ should be a unitary representation of $hs(2,2)$ that
decomposes into massless weight spaces under $SO(4,2)$. This
condition is necessary, provided that the theory has an expansion
around a maximally symmetric AdS vacuum, which must be invariant
under global $hs(2,2)$ valued gauge transformations. Moreover, in
order for $hs(2,2)$ to have a well-defined action on $\cS$ we must
demand that $K=0$ in $\cS$. This shows that $\cS$ must be made up
of tensor products of two spin $j$ doubletons with opposite
eigenvalue of $K$, that is $[D(j,0;j+1)\otimes D(0,j;j+1)]_{\rm
S,A}$.

In order to determine which values of $j$ contribute to $\cS$, we
can study the gauging of $hs(2,2)$ and examine the resulting
curvature constraints (i.e. generalizations of the spin two
Einstein equation) at the linearized level, which would yield
information of the spin $s\geq 2$ sector of $\cS$. Incorporating
the spin $s\leq 1$ sector in an $hs(2,2)$ symmetric fashion
amounts to introducing a scalar master field in some
representation $R$ of $hs(2,2)$. The uncertainty in the choice of
$R$ implies, however, that there is an uncertainty also in the
spin $s\geq 2$ sector, since it is possible that $R$ contains not
just physical spin $s\leq 1$ fields but also physical spin $s\geq
2$ fields. This is in fact the case in supersymmetric extensions
of this bosonic model, as we shall comment on in Section
\ref{sec:sr}. Thus, a determination of $\cS$ based on gauging
alone may have to involve a rather elaborate ansatz, unless one is
willing to accept some loss of generality or one invokes some
other basic principle.

In order to determine $\cS$ we assume that the $hs(2,2)$ gauge
theory is some limit of string theory. Since the gauge theory has
an AdS vacuum, we assume that it describes a bosonic truncation of
the residual Type IIB string bulk dynamics in the near-horizon
region of $N$ coincident three-branes in a decoupling limit in
which the boundary conformal field theory is invariant under
$hs(2,2)$. The precise definition of this limit, which was
suggested recently by \cite{su1,su2}, is discussed further in Sec.
\ref{sec:sr}. We are thus led to imposing the additional
requirement that the spectrum-generating doubleton representations
form a unitary irreducible representation of $hs(2,2)$. From the
above analysis it follows that this uniquely selects the spin
$j=0$ representation $D(0,0;2)$. In the above limit, the bosonic
truncation of the boundary theory is a 4D free scalar in the
adjoint representation of the (global) $SU(N)$ symmetry. The
massless higher spin states emerge in the sector of bilinears in
the scalar field and its derivatives which can be written as
single traces \cite{su1,su2}. For example, the mass-operator gives
rise to a scalar state, while the remaining states corresponds to
a set of higher spin currents \cite{vr2,fz}. Importantly, the
spectrum has no spin one state, as the corresponding current is a
descendant (total derivative) of the mass-operator. This implies
that the massless higher spin spectrum $\cS$ is given by the
symmetric tensor product:

\be \cS=\left[D(0,0;1)\otimes D(0,0;1)\right]_{\rm S}\ .\la{s}\ee
It follows from \eq{stp} that $\cS$ consists of the physical
states of five-dimensional massless fields with spins
$s=0,2,\dots$ and energies $E=s+2$. We remark that the
anti-symmetric part of the tensor product contains states with odd
spins, which from the boundary point of view correspond to states
which are descendants, as explained above in the case of spin one.
We also note that from the point of view of reconstructing the
bulk theory from a boundary theory which has vanishing $K$ using
the holographic correspondence between the generating functional
of boundary correlators and the effective bulk action, it is
perfectly natural that the bulk theory is based on the higher spin
algebra $hs(2,2)$ with the  central $U(1)$ generator $K$ modded
out by an appropriate coset construction.
\footnote{It is also possible to eliminate $K$ from $\cP$ by going
to the Lie algebra of elements obeying $K\star F=0$, $F\in \cP$.
This equation can be solved by an infinite expansion in $K^2$
using elements of the form \eq{x}; see \eq{qar} and below. This
Lie algebra gives rise to the same field content as $hs(2,2)$ upon
gauging, but it seems unnatural from the boundary point of view
and leads to undesirable complications of the algebra as well. We
expect that it can be ruled out at the linearized level; see also
footnote 6. }

%%%%%%%%%%%%%%%%%%%%%%%%%%%%%%%%%%%%%%%%%%%%%%%%%%%%%%%%%%%%%%%%%%%%%

\section{Gauging $hs(2,2)$}\la{sec:g}

%%%%%%%%%%%%%%%%%%%%%%%%%%%%%%%%%%%%%%%%%%%%%%%%%%%%%%%%%%%%%%%%%%%%%

In order to realize $hs(2,2)$ as a local symmetry in a field
theory with spectrum $\cS$ we need to address the following two
basic issues. Firstly, gauging of $hs(2,2)$ introduces both
dynamic gauge fields and auxiliary gauge fields. Fortunately the
structure of a set of gauge fields and curvature constraints that
give rise to one massless spin $s$ degree of freedom are known at
the linearized level in an expansion around AdS spacetime, albeit
in $SO(4,1)$ basis, instead of the spinor basis introduced in the
previous section. Thus in order to give the linearized $hs(2,2)$
valued constraints it suffices to find a one-to-one map between
these two bases.

Secondly, the spectrum $\cS$ in \eq{s} contains a spin zero state.
\footnote{In the $N=4$ supersymmetric case this state becomes the
lowest spin state of a spin four multiplet, as explained in Sec.
6.}
In order to incorporate this degree of freedom while retaining
manifest $hs(2,2)$ gauge invariance, it is natural to generalize
Vasiliev's four-dimensional formulation of higher spin theory and
identify the spin zero mode with the leading component of a scalar
master field $\Phi$ in a particular representation of $hs(2,2)$ to
be identified below. Its remaining components should be the
components of the curvature that are non-vanishing on-shell, that
in the spin two case is known as the Weyl tensor and that are
referred to as the generalized Weyl tensors in the cases of higher
spin, as well as the derivatives of the scalar field and the
generalized Weyl tensors.

In the remainder of this section we are concerned with
establishing the equivalence between the spinorial basis \eq{tx}
for $hs(2,2)$ and the tensorial basis of \cite{vhd1,vhd2} and to
give the definition of the scalar master field. The linearized
analysis is given in the next section.

We begin by introducing the $\cP$ valued one-form

\be A=dx^\m A_\m(y,\yb)\ ,\quad \t_\y(A)=-A\ ,\quad
(A)^{\dagger}=-A\ ,\la{ac}\ee
and a zero-form $B$ in the following representation $\cR$ of
$\cP$:

\be \t_\y(B)=\pi(B)\ ,\quad (B)^{\dagger}=\pi(B)\ . \la{fc}\ee
Here $\pi$ is the linear map

\be \pi(y_\a)=\yb_\a\ ,\quad \pi(\yb_\a)=y_\a\ ,\ee
which acts as an involution of the algebra $\cA$:

\be \pi(F\star G)=\pi(F)\star \pi(G)\ .\la{i}\ee
The $\cP$ gauge transformations are given by:

\be \d_\e A=d\e+[A,\e]_*\ ,\quad \d_\e B=B\star \C(\e)-\e\star B\
,\la{trp}\ee
where $\e$ is a $\cP$ valued local parameter, such that the
following curvature and covariant derivative obey \eq{ac} and
\eq{fc} and are $\cP$ covariant:

\bea F_A&=&dA+A\wedge\star A\ ,\ \quad\qquad\qquad \d_\e
F_A=[F_A,\e]_\star\
,\la{fa}\\
D_A B&=&dB-B\star \pi(A)+A\star B\ ,\quad \d_\e D_A B =D_A B
\star\pi(\e)-\e\star D_A B\ . \la{cd}\eea
To show that $\d_\e B$ and $D_A B$ obey \eq{fc} one needs to use
\eq{dagger}, \eq{ai}, \eq{i}, $\pi^2=1$ and that
$\t_\y(\pi(F))=-\pi(F)$, $(\C(F))^{\dagger}=-\pi(F)$ for $F\in
\cP$. For example, to show that $\d_\e(B)$ obey
$\t_\y(\d_\e(B))=\pi(\d_\e(B))$ we compute:

\be \t_\y(\d_\e B)=\t_\y(B\star \pi(\e)-\e\star
B)=\t_\y(\pi(\e))\star \t_\y(B)-\t_\y(B)\star \t_\y(\e)\nn\ee\be
=-\pi(\e)\star \pi(B)+\pi(B)\star \e=\pi(B\star \pi(\e)-\e\star
B)=\pi(\d_\e B)\ .\ee
We also remark that a similar calculation shows that $B$ indeed
belongs to a representation of $\cP$, that is

\be [\delta_{\e_1},\delta_{\e_2}]B=\d_{[\e_1,\e_2]_*}B\ .\ee

Next we define the $hs(2,2)$-valued gauge field, curvature and
gauge parameter by

\be [A]\ ,\quad F_{[A]}=[F_A]\  ,\quad [\e]\ ,\ee
where we use the notation of \eq{hs}. The gauge transformations
read:

\be \d_{[\e]}[A]=[\d_\e A]\ ,\quad \d_{[\e]}F_{[A]}=[\delta_\e
F_A]\ .\ee
By construction the above expressions are independent of the
choice of representative in $\cP$ of the various $hs(2,2)$-valued
quantities. We note that the curvature and the gauge
transformations are computed by first evaluating the ordinary
$\star$ product \eq{st} between the representatives and then
expanding the result with respect to the particular ordering of
oscillators defined by \eq{tx} and \eq{r} and finally discarding
any terms in $\cI$. In case one would have to perform several
repeated multiplications of objects in $hs(2,2)$ the last step may
of course be carried out at the end, as the operation of modding
out $K$ commutes with taking the $\star$ product.

The $hs(2,2)$-valued gauge field $[A]$ can be represented by a
$\cP^{(0)}$-valued gauge field $A^{(0)}$, which has an expansion
in terms of component fields with tangent indices corresponding to
the traceless multispinors $X^{(0)}$ defined in \eq{tx}, obeying
the reality condition \eq{rc}. The level $\ell$ generators of
$\cP^{(0)}$ thus gives rise to component gauge field with
`spin'
\footnote{We remark that upon linearizing around AdS space,  the
actual AdS spin $(j_L,j_R)$ of the various irreducible components
obtained from a level $\ell$ gauge field take values such that
$j_L+j_R=s,s-2$, while the physical degrees of freedom emerging
from the $\ell$th level has $j_L=j_R=s/2$.}
$s=2\ell+2$, where $s$ is defined to be $1$ plus the internal
spin, which equals $2\ell+1$ in the $\ell$th level.

A real and $C_{\a\b}$ traceless multi-spinor $T_{\a_1\dots
\a_n,\b_1\dots \b_n}$, ($n=2\ell+1=s-1$ at the $\ell$th level of
$\cP^{(0)}$), can be decomposed into irreducible multi-spinors
$T^{(p,m)}_{\a_1\cdots\a_n,\b_1\cdots\b_n}$ with index structures
corresponding to the $SU(2,2)$ Young tableaux:

\be
\begin{picture}(100,100)(50,-20)
\put(-50,20){\makebox(0,0){$T^{(p,m)}_{\a_1\cdots\a_n,\b_1\cdots\b_n}=$}}
\put(0,0){\line(1,0){100}} \put(0,20){\line(1,0){220}}
\put(0,40){\line(1,0){220}} \put(0,0){\line(0,1){40}}
\put(20,0){\line(0,1){40}} \put(80,0){\line(0,1){40}}
\put(100,0){\line(0,1){40}} \put(50,10){\makebox(0,0){$\cdots$}}
\put(50,30){\makebox(0,0){$\cdots$}} \put(120,20){\line(0,1){20}}
\put(140,20){\line(0,1){20}} \put(180,20){\line(0,1){20}}
\put(200,20){\line(0,1){20}} \put(220,20){\line(0,1){20}}
\put(160,30){\makebox(0,0){$\cdots$}}
\put(10,30){\makebox(0,0){$\bullet$}}
\put(90,30){\makebox(0,0){$\bullet$}}
\put(110,30){\makebox(0,0){$\bullet$}}
\put(190,30){\makebox(0,0){$\bullet$}}
\put(50,-10){\makebox(0,0){$\underbrace{\hspace{3.5
cm}}_{\mbox{$p$ boxes}}$}}
\put(160,50){\makebox(0,0){$\overbrace{\hspace{4.2
cm}}^{\mbox{$2m$ boxes}}$}} \put(230,20){\makebox{,\quad $p+m=n$\
,}}
\end{picture}
\la{yt1} \ee
where the `undotted' and `dotted' boxes refer to spinor indices
contracted with $y$ spinors and $\bar{y}$ spinors, respectively.
These spinors belong to equivalent representations of $SU(2,2)$,
and hence their indices can be put in the same Young tableaux.
Since the spinors are Grassmann even two dotted or undotted boxes
cannot be placed on top of each other. To count the (real)
dimension $d_{p,m}$ of the Young tableaux \eq{yt1} we thus first
compute the complex dimension $D_{p,m}$ by performing a
`$SU(4)$-count' in which the dotted and undottedness is neglected.
These correspond to imposing the reality condition (it is not
important whether $T$ is real or purely imaginary), which implies
$d_{p,m}=D_{p,m}$. We remark that the reality condition of course
requires $SU(2,2)$ spinors. Taking also the tracelessness
condition in \eq{tx} into account one finds that the dimension of
\eq{yt1} is given by

\bea d_{p,m}&=&{4\cdot 5\cdots (3+2m+p)\cdot 3\cdot 4 \cdots
(2+p)\over (2m+p+1)\cdot(2m+p)\cdots (2m+1)\cdot
(2m)\cdot(2m-1)\cdots 1\cdot p\cdot (p-1)\cdots
1}\nn\\&&-\mbox{same with $p\rightarrow p-1$}\nn\\
&=&{1\over 12}(p+2)(p+1)(2m+1)(2m+p+2)(2m+p+3)-\mbox{same with
$p\rightarrow p-1$}\nn\\&=&{2\over
3}(m+p+\ft32)(m+\ft12)(2m+p+2)(p+1)\ .\la{d1}\eea
It has been shown \cite{vhd1,vhd2} that linearized curvature
constraints (see Sec.  \ref{sec:c}) leading to the on-shell
massless spin $s$ weight space $D(\ft{s}2,\ft{s}2;s+2)$ of
$SO(4,2)$ can be written using a space of five-dimensional gauge
fields with tangent space indices given by irreducible $SO(4,1)$
tensors $T^{(p,m)}_{a_1b_1,\dots,a_mb_m;c_1\dots c_p}$, $m+p=s-1$,
corresponding to the following Young tableaux:

\be
\begin{picture}(100,100)(45,-20)
\put(-55,20){\makebox(0,0){$T^{(p,m)}_{a_1b_1,\dots,a_mb_m;c_1\dots
c_p}=$}} \put(0,0){\line(1,0){100}} \put(0,20){\line(1,0){200}}
\put(0,40){\line(1,0){200}} \put(0,0){\line(0,1){40}}
\put(20,0){\line(0,1){40}} \put(40,0){\line(0,1){40}}
\put(80,0){\line(0,1){40}} \put(100,0){\line(0,1){40}}
\put(60,10){\makebox(0,0){$\cdots$}}
\put(60,30){\makebox(0,0){$\cdots$}} \put(120,20){\line(0,1){20}}
\put(140,20){\line(0,1){20}} \put(180,20){\line(0,1){20}}
\put(200,20){\line(0,1){20}} \put(160,30){\makebox(0,0){$\cdots$}}
\put(50,-10){\makebox(0,0){$\underbrace{\hspace{3.5
cm}}_{\mbox{$n_2=m$ boxes}}$}}
\put(100,50){\makebox(0,0){$\overbrace{\hspace{7
cm}}^{\mbox{$n_1=m+p=s-1$ boxes}}$}} \put(210,20){\makebox{,\quad
$0\leq n_1\leq n_2$\ ,}}
\end{picture}
\la{yt2} \ee
Here the notation is such that the pair of indices $a_ib_i$
$(i=1,...,m)$ goes into the $i$th pair of anti-symmetrized boxes
and $c_1\dots c_p$ into the remaining $p$ symmetrized boxes. The
irreducible Young tableaux \eq{yt2} has dimension

\be d'_{p,m}=\ft23(n_1+\ft32)(n_2+\ft12)(n_1+n_2+2)(n_1-n_2+1)\
\la{d2}\ee
The dimensions \eq{d1} and \eq{d2} agree for $n=s-1$, so that
\eq{yt2} can be converted into \eq{yt1} by making use of Dirac
matrices. Thus the spin $s$ gauge fields in the spinorial basis
are in one-to-one correspondence with the spin $s$ gauge fields in
the Lorentzian basis of \cite{vhd1,vhd2}. We emphasize, however,
that whereas the latter formulation is only defined in a
linearization around AdS, and thus does not contain any
information about the full higher spin gauge algebra, this data
are naturally incorporated in the spinorial formalism used here.

The master field $B$ can be expanded in terms of elements of the
form

\bea && {1\over(n!)^2}K^{k}B^{(p,m;k)}_{\a_1\dots
\a_n,\b_1\dots\b_n}\yb^{\a_1}\cdots \yb^{\a_n}y^{\b_1}\cdots
y^{\b_n})\equiv K^kB^{(p,m;k)}\ , \la{r}\w2 && C^{\a_1\b_1}
B^{(p,m;k)}_{\a_1\dots \a_n,\b_1\dots\b_n}=0\ , \nn\w2
&&p,m,k\geq 0\ ,\quad p+m=n\ ,\quad m=0,2,4,\ ,\dots \eea
where the superscripts $(p,m)$ refer to the index structure
defined by the Young tableaux \eq{yt1}. Note that \eq{fc} and
$\t_\y(K)=\pi (K)$ implies that
$\t_\y(B^{(p,m;k)})=\pi(B^{(p,m;k)})$, which in turn implies that
$(-1)^{m}B^{(p,m;k)}=B^{(p,m;k)}$, where we used that
$\t_\y(P^a)=\pi(P^a)$ and $\t_\y(M^{ab})=-\pi(M^{ab})$. Thus $m$
must be even while $p$ can be any integer, as indicated in
\eq{r}. From \eq{fc} it follows that the multi-spinors in \eq{r}
obey the following reality condition (note the difference to
\eq{rc}):

\be \bar{B}^{(p,m;k)}_{\a_1\dots \a_n,\b_1\dots
\b_n}=B^{(p,m;k)}_{\a_1\dots \a_n,\b_1\dots \b_n}\ .\ee
Modulo the degeneracy due to $K$, the field content of the master
scalar field therefore falls into `trajectories' $B^{(p,m)}$,
$p=0,1,2,...$. The one-to-one map between \eq{yt1} and \eq{yt2}
shows that the leading component $B^{(0,m)}$ defines a traceless
Lorentzian spin $s=m$ tensor carrying $m$ pairs of anti-symmetric
indices. For $s=2,4,...$ this is exactly the index structure of
the spin $s$ Weyl tensor that was introduced in the linearized
analysis of \cite{vhd1,vhd2}, while $B^{(0,0)}$ is the expected
(real) scalar field. For $p>0$ the index structure $B^{(p,m)}$ is
exactly that of $p$ derivatives of the leading component. Thus,
apart from the degeneracy, the desired field content of the
master scalar field emerges in $B$.

Next, we proceed by defining the master scalar field $\F$ in the
representation $R$ of $hs(2,2)$, where $R$ is the subspace of the
representation space $\cR$ of $\cP$ defined by
\footnote{It is also possible to eliminate $K$ from the master
scalar field by considering an expansion of  $B$ in terms of
elements of the form $(K*)^k*B^{(p,m;k)}$ instead of \eq{r}. This
is analogous to the definition of the basis \eq{tx} for $\cP$.
This leads to a unique decomposition of $\cR=\cR^{(0)}+\cR'$,
where $\cR'=\cR^{(1)}+\cR^{(2)}+\cdots$. The space $\cR'$ is an
$\cP$ invariant subspace. Thus $\cR/\cR'$ is a representation
space for $hs(2,2)$ and we can define the $hs(2,2)$
transformations and the covariant derivative by
$\d_{[\e]}[B]=[\delta_\e B ]$ and $D_{[A]}[B]=[D_A B]$. However,
using these results, a careful analysis of the scalar field
equation shows that while \eq{se1} still holds, the trace term on
the right hand side of \eq{se2} is now absent. This in turn gives
$m^2=-5$, which leads to a scalar field in a non-unitary
representation (complex $E$). This algebraic construction is
therefore pathological.}

\be
K\star \Phi=0\ ,\quad \Phi\in \cR\ .
\la{qar}
\ee
This condition serves two purposes. Firstly, it assures that the
$hs(2,2)$-covariant derivative and $hs(2,2)$-gauge
transformations of $\Phi$, which are given by

\be D_{[A]}\Phi=d\Phi + \Phi\star \pi(A)-A\star \Phi\ ,\quad
\d_{[\e]}\Phi=\Phi\star \pi(\e)-\e\star \Phi\ ,\ee
are well-defined, i.e. independent of the choice of
representative for $[A]$ and $[\e]$ and obeying $K\star
D_{[A]}\Phi=K\star \delta_{[\e]}\Phi=0$. To see this we use \eq{i}
and $\pi(K)=-K$ and the fact that $K$ is central. Secondly,
\eq{qar} removes the degeneracy due to $K$. To see this we first
solve \eq{qar} by using the following lemma:

\be K*\large(K^k T_{(2n)}\large)=K^{k+1}T_{(2n)}-{1\over 4}k(k+2n+3)
K^{k-1}T_{(2n)}\ ,\la{lemma}\ee
where $T_{(2n)}$ is a monomial of rank $2n$ with traceless
multi-spinor coefficients $T_{\a_1\dots \a_n,\b_1\dots \b_n}$ as
follows:

\be T_{(2n)}={1\over (n!)^2}T_{\a_1\dots \a_n,\b_1\dots \b_n}
y^{\a_1}\cdots y^{\a_n}\yb^{\b_1}\cdots \yb^{\b_n}\ ,\quad
C^{\c\d}T_{\c\a_2\dots \a_n,\d\b_2\dots \b_n}=0\ .\ee
We note that in computing \eq{lemma} the single-contractions
cancel while the double-contractions are of three types: those
involving $K^k$, which give a factor of $-{1\over
4}k(k+3)$; the mixed ones, which give $2(-{1\over 4} kn)$; and those
involving $T_{(2n)}$, which are proportional to the
(vanishing) trace. In particular, the tracelessness of a spinor
space element $F$ can be written as $K\star F=KF$. Using
\eq{lemma} the condition \eq{qar} can solved recursively, leading
to the following general solution:

\bea \Phi^{(p,m;2k)}&=&{1\over k![k+p+m+\ft32]_k}\Phi^{(p,m;0)}\ ,\la{even}\\
\Phi^{(p,m;2k+1)}&=&0\ ,\la{odd}\eea
where we use the notation of \eq{r} and the Pochhammer symbol
$[a]_b\equiv a(a-1)\cdots (a-b+1)$, for $b$ positive integer.
Thus the degeneracy is completely removed. Each independent
(traceless) tensorial structure $\Phi^{(p,m)}\equiv
\Phi^{(p,m;0)}$ gives rise to an infinite expansion in terms of
even powers of $K$, such that $\Phi$ can be written in terms of
elements of the form

\be f(p+m;K^2)\Phi^{(p,m)}\ ,\la{kexp}\ee
where the analytic function function $f(n;z)$ is defined by

\be f(n;z)=\sum_{k=0}^{\infty} {z^k\over k!(k+n+\ft32)_k}\ .\ee
For example, the scalar field is represented by the expansion:

\be (1+{2\over 5}K^2+{2\over 35}K^4+{4\over 945}K^3+\cdots)\f\
,\la{ske}\ee
where $\phi$ is the $y$ and $\yb$ independent component of
$\Phi$.

%%%%%%%%%%%%%%%%%%%%%%%%%%%%%%%%%%%%%%%%%%%%%%%%%%%%%%%%%%%%%%%%%%%%%

\section{Linearized constraints }\la{sec:c}

%%%%%%%%%%%%%%%%%%%%%%%%%%%%%%%%%%%%%%%%%%%%%%%%%%%%%%%%%%%%%%%%%%%%%

The first step towards finding the full field equations for the
higher spin gauge theory based on the $hs(2,2)$ algebra is to
identify the appropriate linearized field equations. The requisite
for writing these are the $hs(2,2)$ covariant curvature and scalar
master fields defined in the previous section. The basic
assumption is that the higher spin gauge theory should make sense
as an expansion around the AdS vacuum described by

\be \F=0\ ,\quad [A]=[\Omega]\ ,\ee
where $\Omega$ is the `flat' AdS connection\footnote{We have
chosen units such that the AdS radius $R_{AdS}=1$. It can be
introduced by replacing $P_a\rightarrow R_{AdS}P_a$. The
insertions of powers of $R_{AdS}$ in the component formulae are
then determined by dimensional analysis.}:

\be \Omega_\m=i( e_\m{}^aP_a+\ft12 \omega_\m{}^{ab}M_{ab})\
,\quad F_\Omega=  d\Omega+\Omega\star\Omega= i\large(T^aP_a+\ft12
(R^{ab}+e^a\wedge e^b)M_{ab}\large) =0\ . \la{omega}\ee
Here $e_\m{}^a$ and $\omega_\m{}^{ab}$ are the f\"unfbein and
Lorentz connection and $T^a$ and $R^{ab}$ the torsion and Riemann
curvature two-forms defined by:

\be T^a=de^a+\omega^a{}_b\wedge e^b\ ,\quad
R^{ab}=d\omega^{ab}+\omega^a{}_c\wedge \omega^{cb}\ .\ee
The resulting five-dimensional Einstein equation with cosmological
constant $\Lambda$ reads:

\be R_{\m\n}-{1\over 2}(R +\Lambda)g_{\m\n}=0\ ,\quad
\Lambda=-{12\over R^2}\ ,\la{ee}\ee
where the metric and the Ricci tensor have been defined by

\be g_{\m\n}=e_\m{}^ae_{\n a}\ ,\quad R_{\n a} =e_b{}^\m
R_{\m\n,a}{}^b\ .\ee
In the AdS vacuum we find

\be R_{\m\n}{}^{ab}=-2e_{[\m}{}^ae_{\n]}{}^b\ ,\quad R_{\m\n}=4
g_{\m\n}\ .\ee
The normalization is such that the AdS metric is given by

\be ds^2={1\over r^2}(dr^2+dx^2)\ ,\la{adsm}\ee
in five-dimensional Poincar\'e coordinates.

Assuming that the full equations have a curvature expansion in
powers of $\F$,  a linearization of these curvatures around the
AdS background should give rise to free equations describing the
massless degrees of freedom in the spectrum $\cS$ defined in
\eq{s}, that is, the  free equations for massless fields of spin
$s=0,2,4,...$ and AdS energy $E=s+2$. In the case of $s=0$ this
corresponds to the Klein-Gordon equation
\footnote{Using \eq{adsm} and making the ansatz $\f\sim r^{E}$ we
find that the Klein-Gordon equation
$(\nabla^\m\nabla_\m-m^2)\f=0$, which follows from the usual free
action $\ft12\int
d^5x\sqrt{-g}(\partial^\m\f\partial_\m\f-m^2\f^2)$ with `positive'
$m^2$, leads to the characteristic equation $E(E-4)=m^2$, where
$E$ and $m^2$ is given in units of $R_{AdS}$. For $E=2$ we find
$m^2=-4$, which saturates the lower bound for $m^2$ (see, for
example, \cite{fz}). }:

\be (\nabla^\m\partial_\m+4)\f=0\ ,\la{sfe}\ee
where $\f$ is an independent scalar (which will turn out to be the
leading component of the master scalar field). The linearized spin
two equation can of course be obtained by linearizing \eq{ee}.
However, the formalism that appears to be the most convenient in
the context of higher spin gauge theory is a generalization of the
first order constraint formulation of \eq{ee}. In the spin two
case this amounts to solving for the auxiliary Lorentz connection
in terms of the dynamical f\"unfbein from the torsion constraint
$T^a=0$, and writing the Einstein equation as a constraint on the
AdS covariantization $F^{ab}=R^{ab}+e^a\wedge e^b$ of the Riemann
curvature. This tensor contains $50$ components in
five-dimensions, of which $15$ are set equal to zero by the
Einstein equation. The remaining $35$ non-vanishing components
define the spin two Weyl tensor. It corresponds to the Young
tableaux \eq{yt2} with $m=2$ and $p=0$ (the `window' diagram), or
equivalently a multi-spinor with Young tableaux \eq{yt1} defining
a totally symmetric multi-spinor $\F^{(0,2)}_{\a\b\c\d}$. This
choice of $m$ and $p$ does not correspond to an algebra element;
the Weyl tensor is obtained by converting both the algebra-valued
tangent space indices $ab$ and the curved indices $\m\n$ on the
spin two curvature $F_{\m\n,ab}$ into spinor indices by using the
f\"unfbein and Dirac matrices. Thus, in this language the (full)
Einstein equation with cosmological constant \eq{ee} can be
written as the following constraint:

\be F_{\m\n,ab}=(\C_{ab})^{\a\b}(\C_{cd})^{\c\d}e_\m{}^c e_\n{}^d
\F^{(0,2)}_{\a\b\c\d}\ ,\quad T_{\m\n,a}=0\ .\ee

The higher spin generalization of these curvature constraints has
been given in the free case in a linearization around the AdS
vacuum in \cite{vhd1,vhd2} using a tensorial basis with Lorentzian
indices. These constraints are straightforward to cast into the
spinorial basis. The higher spin dynamics also requires a
constraint on the scalar master field $\F$. Since it is already
linear in fluctuations, the only possible constraint linearized
constraint on $\F$ is the vanishing of $D_\O\F$. Using the
notation of \eq{yt1} the linearized constraints therefore read
($n=2\ell+1$)

\bea
&& F^{(0)}_{\a_1\dots \a_n,\b_1\dots\b_n}= e^a\wedge e^b
(\C_{ab})^{\c\d}\F^{(0,n+1)}_{\c\a_1\dots \a_n,\d\b_1\dots\b_n}\
,\la{t}
\\[15pt]
&& d\F+\Omega\star \F-\F\star
\pi(\Omega)= 0\ ,
\la{f}
\eea
where $F^{(0)}$ is the linearized curvature

\be F^{(0)}=dA^{(0)}+\Omega\star A^{(0)}+A^{(0)}\star\Omega \
,\la{f0}\ee
Here the superscript $(0)$ refers to our representing $hs(2,2)$ by
$\cP^{(0)}$ ($A^{(0)}$, $F^{(0)}$ and $\F$ are all assumed to be
fluctuations around the AdS vacuum). At the linearized level, the
last two terms in \eq{f} are manifestly traceless; commutation
with the AdS connection does not give rise to any terms in the
ideal $\cI$. The left-hand side of \eq{t} contains all possible
spinorial index structures compatible with the fact that $F^{(0)}$
is an element of the $\ell$th level of $\cP^{(0)}$, while the
right-hand side only contains the symmetric spin $s=n+1$ tensor
$\F^{(0,n+1)}$ (without $K^2$-expansion), which is the higher spin
generalization of the spin two Weyl tensor
$\F^{(0,2)}_{\a\b\c\d}$. Thus \eq{t} contains generalized torsion
constraints, field equations as well as the identification of the
generalized Weyl tensors.

We remark that whereas the constraint \eq{f} on the master scalar
field is written in terms of functions of $y$ and $\yb$, the
constraint \eq{t} on the curvature has been written in component
form. The reason for this is that whereas the full constraint on
the master scalar field has to be of the form $D_A\F={\cal
V}_1(\F)$, where ${\cal V}$ is linear
\footnote{It is not obvious
that ${\cal V}_1$ cannot depend on $d\F$; in fact, in four
dimensions this fact was shown only recently in \cite{ss3}. We
expect that the same holds in five dimensions.}
in $A$ and quadratic in $\F$, so that its linearization is given
uniquely by \eq{f}, the full form of the curvature constraint
\eq{t} is $F_A={\cal V}_2(A;\F)$ for some function ${\cal V}_2$
which is quadratic in $A$ and linear in $\F$. Thus the implication
of \eq{t} is that whatever form ${\cal V}_2$ has, its
linearization around the AdS vacuum must be given by the
right-hand side of \eq{t}. Some further remarks on the curvature
expansion of the full theory are given in the Conclusions.

The constraints \eq{t} and \eq{f} are integrable. The
integrability of \eq{f} follows from the flatness of $\O$, as
given in \eq{omega}. The integrability of \eq{t} requires the
Bianchi identity

\be dF^{(0)}+\Omega\star F^{(0)}-F^{(0)}\star\Omega= 0\la{bi}\ee
to be satisfied when $F^{(0)}$ is substituted using \eq{t}. To
examine this equation we write the constraints \eq{t} and \eq{f}
in component form:

\bea F^{(0)}_{\mu\nu,\a_1\dots \a_n,\b_1\dots \b_n}&\equiv&
2\nabla_{[\m}A^{(0)}_{\n],\a_1\dots \a_n,\b_1\dots \b_n}\nn\\&&+
n\left((\C_{[\m})^{\phantom{(0)}}_{(\b_1|}{}^{\c}A^{(0)}_{\n],\a_1\dots
\a_n,\c|\b_2\dots \b_n)}-
(\C_{[\m})^{\phantom{(0)}}_{(\a_1|}{}^{\c}A^{(0)}_{\n],\c|\a_2\dots \a_n),\b_1\dots \b_n}\right)\nn\\
&=&{1\over 8}(\C_{\mu\nu})^{\c\d}\F^{(0,n+1)}_{\c\a_1\dots
\a_n,\d\b_1\dots \b_n}\ , \la{t2}\eea \be
\nabla_\m\F_{\a_1\dots\a_n,}{}^{\b_1\dots\b_n}- {1\over 2}
(\C_\m)^{\c\d}\F_{\c\a_1\dots\a_n,\d}{}^{\b_1\dots\b_n}+
{n^2\over 2}(\C_\m)_{(\a_1}{}^{(\b_1}
\F_{\a_2\dots\a_n),}{}^{\b_2\dots\b_n)}=0\ .\la{f2}\ee
Here $\nabla_\m$ is the Lorentz covariant derivative. We note that
the multi-spinors in the last equation are the coefficients of
the $y$ and $\yb$ expansion of the master scalar field including
the $K^2$-expansions \eq{kexp}. The component form of the Bianchi
identity \eq{bi} reads:

\be \nabla_{[\m}F^{(0)}_{\n\r],\a(n),\b(n)}+{n\over 2}
\left((\c_{[\m})_{\b}{}^{\c}F^{(0)}_{\n\r],\a(n),\c\b(n-1)}-
(\c_{[\m})_{\a}{}^{\c}F^{(0)}_{\n\r],\c\a(n-1),\b(n)}\right)=0\
.\la{bi2} \ee
By inserting \eq{t2} in \eq{bi} and making use of \eq{f2} to
substitute for $\nabla_\m\F^{(0,m)}$ by

\be
\nabla_\m\F^{(0,m)}_{\a_1\dots \a_m,\b_1\dots\b_m}=
\ft12(\C_\m)^{\c\d}\F^{(1,m)}_{\c(\a_1\dots\a_m|,\d|\b_1\dots\b_m)}\ ,
\ee
where $(0,m)$ and $(1,m)$ refers to the index structure according
to \eq{yt1} (the $K$-expansion of $\Phi$ does not affect this
sector), it follows that \eq{bi} holds due to the following Fierz
identities
\footnote{The spinor conventions given in Sec. 2 are
such that the following Fierz identity holds: \be
M^{\a\b}N^{\c\d}=-{1\over8}(M\C^{ab}N)^{\a\d}(\C_{ab})^{\b\c}-
{1\over
4}(M\C^aN)^{\a\d}(\C_a)^{\b\c}-{1\over4}(MN)^{\a\d}C^{\b\c}\
.\ee}:

\bea
(\C_{a[b})^{\a\b}(\C_{cd]})^{\c\d}\F^{(0,m)}_{\a\b\c\d\cdots}&=&0\
,\la{fid1}\\
(\C_{[b})^{\a\b}(\C_{cd]})^{\c\d}\F^{(1,m;0)}_{\a\b\c\d\cdots}&=&0\
, \la{fid2}\eea
The symmetries and the tracelessness of the multi-spinors
contracting the Dirac matrices are important for these identities
to be satisfied.

Using the equivalence between the spinorial and tensorial bases
established in Sec. \ref{sec:g} it is straightforward to see that
the constraint \eq{t} is equivalent to the curvature constraints
which were shown in \cite{vhd1,vhd2} to give rise to a massless
spin $s$ degree of freedom. Thus \eq{t} sets all components of the
curvature except the generalized Weyl tensors equal to zero. The
vanishing curvatures are the generalized torsion equations and the
spin $s\geq 2$ field equations. The torsion equations are
algebraic equations for the auxiliary gauge fields
$A^{(p,m;0)}_\m$ with $m+p=2\ell+1$, $m>0$, which can be solved in
terms of the generalized f\"unfbeins $A^{(p,0;0)}_\m$ with
$p=2\ell+1$. The remaining vanishing curvatures then become
second-order field equations, which after gauge fixing give rise
to mode expansions based on the massless $SO(4,2)$ weight spaces
$D(s,s;s+2)$ with $s=2\ell+2$. Thus the gauge fields give rise to
the spin $s\geq 2$ sector of the spectrum \eq{s}.

The non-vanishing curvature components in \eq{t} are those
corresponding to the $SU(2,2)$ Young tableaux \eq{yt1} with
$m=2\ell+2$, $p=0$, that is the $SO(4,1)$ Young tableaux \eq{yt2}
with $n_1=n_2=m$. These are the generalized Weyl tensors
$\F^{(0,m)}$, which are totally symmetric multi-spinors. From the
constraint \eq{f}, which is written in components in \eq{f2}, it
follows that the trajectory $\F^{(p,m)}$ ($p=0,1,2,...$) with
fixed $m$ corresponds to the derivatives of the leading tensor
$\F^{(0,m)}$. Hence the only independent component of the scalar
master field is the single real scalar field $\f\equiv
\F^{(0,0)}|_{y=0}$. From \eq{f2} it follows that

\bea
\partial_\m\f&=&{1\over 2}
(\C_\m)^{\a\b}\F^{(1,0)}_{\a\b}\ ,
\la{se1}\w2
\nabla_\m \F^{(1,0)}_{\a\b}&=& {1\over 2}
(\C_\m)^{\c\d}\Big[ \F^{(2,0)}_{\a\b\c\d}+\ft25
C_{(\a|\b}C_{|\c)\d}\f\,\Big]-{1\over 2}(\C_\m)_{\a\b}\f\ ,
\la{se2}
\eea
where the superscripts $(p,m)$ of the traceless multispinors refer
to the index structure according to \eq{yt1}. The trace part in
the first term on the right-hand side of \eq{se2} comes from the
$K^2$-expansion of the scalar field according to \eq{ske}. This
term is necessary for obtaining the scalar field equation with the
critical mass term that is appropriate for its being AdS-massless,
and hence the importance of the condition \eq{qar}. Indeed,
combining the two equations given above and making use of the
Fierz identity

\be (\C^a)^{\a\b}(\C_a)^{\c\d}\F^{(2,0)}_{\a\b\c\d}=0\ ,\ee
we find that the scalar field satisfies the scalar field equation
\eq{sfe}, which gives rise to a mode expansion based on the spin
zero weight space $D(0,0;2)$ in the spectrum \eq{s}.

%%%%%%%%%%%%%%%%%%%%%%%%%%%%%%%%%%%%%%%%%%%%%%%%%%%%%%%%%%%%%%%%%%%%%

\section{Summary and Remarks}\la{sec:sr}

%%%%%%%%%%%%%%%%%%%%%%%%%%%%%%%%%%%%%%%%%%%%%%%%%%%%%%%%%%%%%%%%%%%%%

We have used Grassmann even spinor oscillators to construct a
bosonic higher spin extension $hs(2,2)$ of the five-dimensional
AdS algebra $SU(2,2)$ containing generators giving rise to
dynamical as well as auxiliary gauge fields with spins
$s=2,4,6,...$ upon gauging. The higher spin algebra is naturally
embedded into a larger algebra $\cP$ as the coset $\cP/\cI$ where
$\cI$ is an ideal of $\cP$ generated by arbitrary
$\star$-polynomials of the central element $K$ multiplied by
traceless polynomials of $y$ and $\yb$ (see \eq{tx} and \eq{ge}).
The large algebra $\cP$ can be represented unitarily and
irreducibly on a spin $j$ doubleton with $|K|={j\over 2}$, while
the higher spin algebra $hs(2,2)$ has a well-defined
representation only on the spin zero doubleton with $K=0$. The
symmetric tensor product of two spin zero doubletons gives rise to
a unitary irreducible representation $\cS$ of $hs(2,2)$ that
decomposes into spin $s=0,2,4,...$ weight spaces of $SO(4,2)$.

We expect $\cS$ to be the spectrum of a five-dimensional gauge
theory with local $hs(2,2)$ symmetry and an AdS vacuum with
unbroken global $hs(2,2)$ symmetry. As a first step towards
constructing this theory, we have shown that the spin $s$ gauge
fields which arises upon gauging $hs(2,2)$ are in one-to-one
correspondence with the set of spin $s$ gauge fields which were
used in \cite{vhd1,vhd2} to construct linearized curvature
constraints describing a massless spin $s$ field in five
dimensions. We have converted these constraints, which were
originally given in the Lorentzian basis, into the spinor basis,
where they can be written as \eq{t}. Furthermore have identified a
representation for the scalar master field which contains the
physical spin zero field, the generalized higher spin Weyl tensors
and their derivatives. The scalar master constraint is simply
given by the vanishing its AdS covariant derivative. We remark
that while the expression \eq{f0} of the linearized curvature is
equivalent to the one used in the formulation of \cite{vhd1,vhd2},
their formulation neither incorporates the oscillator algebra
\eq{st} and \eq{l} required for constructing the expressions
\eq{fa} and \eq{cd} for the non-linear curvature and covariant
derivative, nor the master scalar field $\phi$ constructed in Sec.
4, which plays a crucial role in higher spin gauge theory.

In terms of the oscillator $y_\a$ and its Majorana  conjugate
$\yb_\a$ the $U(2,2)$ subalgebra is spanned by the bilinears
$y_\a\yb_\b$. The remaining generators of $\cP$ form levels
labeled by an integer $\ell$, such that $U(2,2)$ is the zeroth
level and the $\ell$th level is spanned by monomials that contain
$2\ell+1$ $y_\a$-oscillators and the same number of
$\yb_\a$-oscillators. The algebra elements may be written either
using the Weyl-ordered (fully symmetrized) oscillator product as
in \eq{x}, or by extracting explicitly positive powers of $K\star$
as in \eq{tx}. These correspond to traces taken using the
anti-symmetric charge conjugation matrix $C_{\a\b}$. In the latter
basis, the ideal $\cI$ is given by the space of arbitrary
polynomials containing a strictly positive number of $K\star$
factors. Thus $hs(2,2)$ is isomorphic to a space of traceless and
real multispinors that is arranged into levels such that the
elements in the $\ell$th level carry two sets of $2\ell+1$
symmetrized spinor indices (one set contracted with $y$'s and the
other set contracted with $\yb$'s). The $\ell$th level can be
decomposed further by anti-symmetrizing $p$ pairs of spinor
indices taken from the two sets ($0\leq p\leq 2\ell+1$) and
symmetrizing the remaining indices, as represented by the Young
tableaux \eq{yt1}. For given $p$ these are in one-to-one
correspondence with the Lorentz-tensor represented by a two-row
Young tableaux with $2\ell+1$ boxes in the first row and
$2\ell+1-p$ in the second, as given in \eq{yt2}.

Upon gauging, we thus find the gauge field content required for
writing the above-mentioned curvature constraints. The gauge
fields corresponding to the generators with $p=0,...,2\ell$ are
auxiliary, while $p=2\ell+1$ corresponds to the dynamical gauge
field $A_{\m,a_1...a_{2\ell+1}}$. Of particular importance is also
the curvature corresponding to $p=0$, that is
$R_{\m\n,a_1b_1,...,a_{2\ell+1}b_{2\ell+1}}$ where each pair
$a_ib_i$ is anti-symmetric. This curvature contains the only spin
$2\ell+2$ curvature components that are non-vanishing on-shell.
These define the generalized spin $2\ell+1$ Weyl tensor, which is
a fully symmetric, real and traceless multispinor with
$2(2\ell+2)$ spinor indices occurring on the right hand side of
the curvature constraint \eq{t}.

Whereas the gauge fields fit naturally into the adjoint
representation of $hs(2,2)$, perhaps a less obvious issue in the
construction is to determine which $hs(2,2)$ representation
contains the Weyl tensors. To this end, we first observe that
although the Weyl tensors have spins $2,4,...$ it is natural to
fit them into a scalar master field. This is because the
constraint algebra is written as a free differential algebra, or a
Cartan integrable system, which means that for each $p$-form with
$p>0$ there will be a corresponding $(p-1)$-form gauge parameter
(the spacetime diffeomorphism group is automatically incorporated
into the gauge group such that a vector field corresponds to field
dependent $(p-1)$-form gauge parameters given by the inner
derivatives of the corresponding $p$-form potentials). Moreover,
since the spectrum $\cS$ contains a spin zero degree of freedom,
it is natural to attempt to unify the corresponding scalar field
with the Weyl tensors in a scalar master field.

This stage of the construction reveals an intimate interplay
between the group theoretical constraints and the dynamics. The
$hs(2,2)$ transformation property \eq{trp} of the scalar master
field is determined by the requirement that it should contain the
Weyl tensors and the scalar field, which amounts to the constraint
\eq{fc} involving the involution $\pi$. This in turn determines
the form of its gauge covariant derivative \eq{cd}, where we in
particular note the twisting of the connection in the last term by
the insertion of $\pi$. At the linearized level the only natural,
gauge invariant constraint on the scalar master field, which we
treat as a linear fluctuation around a zero background value, is
to set its background covariant derivative to zero. This turns out
to yield the correct scalar equation as well as constraints on the
remaining components of the scalar master field which are
consistent with identifying the fully symmetric higher spin
multispinors with the Weyl tensors (the latter amounts to
verifying the Fierz identities \eqs{fid1}{fid2}). The twisting,
which flips the sign of the f\"unfbein contribution while it keeps
the sign of the contribution from the Lorentz connection, that is
$\pi(P_a)=-P_a$ and $\pi(M_{ab})=M_{ab}$, plays a crucial role in
all this. In fact, if one takes a scalar master field in the
adjoint representation and set its adjoint covariant derivative to
zero, then it will be constant.

It is interesting to note that the definition of $\pi$ in five
dimensions relies on the fact that the Dirac matrices with one and
two vector indices have different symmetry properties. On the
other hand, these Dirac matrices have the same symmetry seven
dimensions. Thus the five-dimensional $\pi$ cannot be generalized
to seven dimensions. The problem of finding appropriate
twist-operations in higher dimensions has been studied further in
\cite{sv}.

Clearly, the analysis in this paper only contains the first step
towards building a full higher spin gauge theory in five
dimensions, and it still remains to construct the interactions. To
this end, we believe that the results of this paper provide the
correct framework for building the interactions. Moreover,
experience with the  higher spin gauge theory in $D=4$ suggests an
efficient method for gauging based on the spinorial formulation
presented here, consisting of embedding of the full, non-linear
constraint algebra into an enlarged constraint algebra based on an
extension of the ordinary spacetime by an auxiliary
non-commutative spinorial $Z$-space a l\'a Vasiliev \cite{v7}.
Indeed suggestions for how this might be done in the case of even
spacetime dimension has already been given quite some time ago
\cite{v8}. We are currently investigating constructions of similar
type in the case of five dimensions, though our results are not
conclusive at this point mainly due to problems with identifying
the proper constraint on the master curvature in the extended
space.

The spinorial oscillators are also useful in constructing
supersymmetric extensions. The higher spin extension $hs(2,2|n)$
of the finite-dimensional supergroup $SU(2,2|n)$ containing the
bosonic subgroup $SO(4,2)\times SU(n)$ and odd supercharges
$Q_\a^i$, $i=1,...,n$, can then be constructed by introducing an
additional set of Grassmann odd complex oscillators $\theta^i$
forming a Clifford algebra, and setting $Q_\a^i=y_\a\theta^i$. The
oscillator realization of $SU(2,2|n)$ contains the generator
$Z=K+\ft12 \theta^i\theta_i$ which becomes central in the higher
spin superalgebra \cite{ss4}. Similar constructions in $D=2n$
involving Kleinian operators have been suggested in \cite{v8}). We
expect the spectrum to be generated by the CPT self-conjugate
superdoubleton, which has vanishing $Z$. Indeed, a preliminary
analysis indicates that the results of this paper will generalize
in a rather straightforward fashion to the maximal case
$hs(2,2|4)$. The main subtlety resides in the fact that the scalar
master field contains not just the gauge matter sector, but also
the spin one three-form field strength and a tower of higher spin
generalizations thereof.

As pointed out recently by \cite{su1,su2} the five-dimensional
sphere compactification of Type IIB string theory with $N\geq1$
units of RR five-form flux and zero string coupling should have a
description in terms of a five-dimensional theory governed by
massless higher spin gauge invariance, which is dual to free
$U(N)$ Yang-Mills theory at the boundary, that is, the theory of
$N^2$ spin one conformal superdoubletons. Zero string/gauge
coupling implies infinite string length (the ratio of the string
length to the radius diverges when the string coupling becomes
small at fixed $N$), which means tensionless strings. The theory
is thus parameterized by the five-dimensional Planck scale
$\ell_p$ and the AdS radius $R=N^{1/3}\ell_p$. The boundary
correlation functions can be constructed from the basic
single-trace operators. In particular the bilinear single-trace
operators give rise to currents that couple to massless AdS modes.
The spectrum of such currents is isomorphic to the product of two
superdoubletons, i.e. the massless spectrum of the above-mentioned
$hs(2,2|4)$ gauge theory.

In the field theory limit $\ell_p<<R$, that is the large $N$
limit, the $hs(2,2|4)$ higher spin gauge theory has a well-defined
curvature expansion \cite{v6,ss3} at energies corresponding to
length scales $\ell$ in the interval $\ell_p<<\ell<<R$. The
above-mentioned duality therefore implies that this expansion is
dual to the interactions of the bilinear currents. Hence this
setup offers a parameter regime in which the strong version of the
Maldacena conjecture can actually be tested directly! Importantly,
even though the 't Hooft coupling vanishes, so that correlation
functions where all operators are single (linear) doubleton fields
are free, the correlation functions involving the
current-bilinears have a non-trivial generating functional, which
should be equal to the effective action of the $hs(2,2|4)$ gauge
theory. Since these interactions persist at zero string coupling
they may be considered to be the basic `M-interactions' defining
M-theory in an unbroken phase.

The boundary theory also contains operators in the form of
normal-ordered products of three or more doubletons. These
correspond to massive bulk modes, which form massive higher spin
multiplets. It would be interesting to investigate whether the
full bulk spectrum originates from a massless higher spin gauge
theory in ten dimensions. Indeed each higher spin multiplet (the
massive ones as well as the massless one) contain a CPT
self-conjugate spin two multiplet. These give rise to a tower of
spin two multiplets describing the Kaluza-Klein modes of the
ten-dimensional supergravity multiplet.

The addition of Yang-Mills interactions break the higher spin
currents in four dimensions \cite{anselmi}. This implies  to that
the bulk theory has a finite string coupling, that is, a finite
string mass. In this massive phase we expect some of the higher
spin gauge symmetries to be realized as St\"uckelberg-like shift
symmetries, with a smooth limit (in the sense that there is no
jump in degrees of freedom) to higher spin gauge symmetry as the
mass-parameter is sent to zero. A better understanding of this
may cast light on the nature of perturbative string theory in AdS
backgrounds as well as on the issue of how to incorporate the
massive multiplets, as some of these may have to be included in
the perturbative spectrum in order for the Higgs mechanism to work
consistently.

We expect the bosonic theory considered in this paper to be a
consistent truncation of the $N=4$ supersymmetric case as follows.
In the boundary we set all fields in a given superdoubleton
multiplet equal to zero except one of the scalars. In the bulk,
the spectrum of the supersymmetric theory consists of a tower of
supermultiplets arranged into levels $\ell=0,1,2...$. In the
truncation to our model, we keep the graviton of the supergravity
multiplet at $\ell=0$, the spin zero and spin four fields at
$\ell=1$ and the field with maximal spin $s_{max}=2\ell+2$ at
level $\ell$. Effectively this amounts to setting $\theta^i=0$ in
the notation introduced above. The bosonic model may therefore
serve as a simplified setup for addressing some of the above
issues, such as the couplings to massive multiplets and the
ten-dimensional origin. The bulk dilaton is thrown away in this
truncation, however, which leaves in doubt whether it may
facilitate the massive string deformation. Further evidence
against this is that the deformation of the boundary scalar theory
by adding a $\f^4$ coupling, which is analogous to the
introduction of finite $g^2_{\rm YM}$ in the super case, breaks
the conformal invariance at the quantum level.

In \cite{ss1} it was conjectured that the seven-sphere
compactification of M-theory with $N$ units of four-form flux
leads to a duality in four spacetime dimensions, which is similar
to the one discussed above in five dimensions. Here the free
three-dimensional singleton is dual to the strong coupling limit,
that is $\ell_p\sim R$, of the four-dimensional higher spin theory
with gauge group $shs(8|4)$. At weak coupling, that is
$\ell_p<<R$, it has a curvature expansion for energies
corresponding to length scales $\ell_p<<\ell<<R$, which is
expected to be dual to the mysterious theory of $N>>1$ coinciding
membranes (and analogously coinciding five-branes are expected to
be dual to weakly coupled higher spin theory in seven dimensions).

In fact, from the higher spin point of view there appears to be a
parity between the IIB and the IIA/11D corners of M-theory, in the
sense that both give rise to similar higher spin gauge theories.
The differences due the presence of string coupling in IIB and the
absence thereof in $D=11$ instead seems to reside with the
patterns of symmetry breaking, which appears to be a property of
the gauged supergravities rather than the full higher spin theory.
Thus, it is tempting to think of an unbroken M-gauge theory
embracing both IIB and 11D in a unified framework. We will
elaborate further on this theme in a separate publication
\cite{ss4}.

To conclude, we believe that higher spin gauge theories do fit
naturally into the M-theory jig-saw and that they will eventually
provide new and fascinating insights to hitherto uncharted limits
of M-theory.

\vspace{5pt}

%%%%%%%%%%%%%%%%%%%%%%%%%%%%%%%%%%%%%%%%%%%%%%%%%%%%%%%%%%%%%%%%%%%%%%%

 \noindent{\Large \bf Acknowledgements}

%%%%%%%%%%%%%%%%%%%%%%%%%%%%%%%%%%%%%%%%%%%%%%%%%%%%%%%%%%%%%%%%%%%%%%%

We are thankful to I. Bars, M. G\"unaydin, D. Roest, J.P. van der
Schaar and E. Witten for stimulating and helpful discussions. This
research project has been supported in part by NSF Grant
PHY-0070964. The work of P.S. is part of the research program of
Stichting voor Fundamenteel Onderzoek der Materie (Stichting FOM).

\pagebreak

%%%%%%%%%%%%%%%%%%%%%%%%%%%%%%%%%%%%%%%%%%%%%%%%%%%%%%%%%%%%%%%%%%%%%%

\end{document}